\documentclass[12pt]{article}
\usepackage{amssymb,amsmath}

\usepackage{curves}
\usepackage{epic}
\usepackage{eepic}
\usepackage{epsfig}

\newcommand{\dd}{\mbox{\rm d}}
\newcommand{\wg}{\wedge}
\newcommand{\gam}{\gamma}

\newcommand{\nth}{\theta}
\newcommand{\dg}{\dagger}
\newcommand{\ddg}{\ddagger}
\newcommand{\tl}{\tilde}

\newcommand{\p}{\partial}

\newcommand{\be}{\begin{equation}}
\newcommand{\bear}{\begin{eqnarray}}

\newcommand{\ear}{\end{eqnarray}}
\newcommand{\ee}{\end{equation}}
\newcommand{\lbl}{\label}
\newcommand{\bi}{\bibitem}
\newcommand{\ci}{\cite}

\newcommand{\vs}{\vspace}
\newcommand{\hs}{\hspace}

\begin{document}

\begin{center}

\
\vs{1cm}

{\bf \Large A Theory of Quantized Fields Based on \\

\vs{2mm}

Orthogonal and Symplectic Clifford Algebras
}

\vs{2mm}

Matej Pav\v si\v c

Jo\v zef Stefan Institute, Jamova 39, SI-1000, Ljubljana, Slovenia; 

email: matej.pavsic@ijs.si
\end{center}

\vs{2mm}

\baselineskip .4cm

\centerline{\bf Abstract}

\vs{2mm}

{\small
The transition from a classical to quantum theory is investigated
within the context of orthogonal and symplectic Clifford algebras, first
for particles, and then for fields. It is shown that
the generators of Clifford algebras have the role of quantum mechanical
operators that satisfy the Heisenberg equations of motion. For quadratic
Hamiltonians, the latter
equations are obtained from the classical equations of motion, rewritten
in terms of the phase space coordinates and the corresponding basis
vectors. Then, assuming that such equations hold for arbitrary path,
i.e., that coordinates and momenta are undetermined, we arrive at the
equations that contains basis vectors and their time
derivatives only. According to this approach, quantization of a classical
theory, formulated in  phase space, is replacement of
the phase space variables with the corresponding basis vectors (operators).
The basis vectors, transformed into the Witt basis,
satisfy the bosonic or fermionic (anti)commutation relations, and can create
spinor states of all minimal left ideals of the corresponding Clifford 
algebra. We consider some specific actions for point particles and fields,
formulated in terms of commuting and/or anticommuting phase space variables,
together with the corresponding symplectic or orthogonal basis vectors.
Finally we discuss why such approach could be useful for grand unification
and quantum gravity.
}

\baselineskip .55cm

\section{Introduction}

During last few decades we are faced with the persisting problems of quantum
gravity and the unification of fundamental interactions. The situation reveals
the need to reformulate the conceptual foundations of physics and to employ a more
evolved mathematical formalism. It has turned out that Clifford algebras provide
very promising tools for description\,\ci{Clifford} and generalization of geometry
and physics\,\ci{CastroHint,PavsicBook,CastroPavsicReview}. There exist two kinds
of Clifford algebras, orthogonal and
symplectic\,\ci{Crumeyrole}. In {\it orthogonal Clifford algebras},
the symmetric product of two basis vectors $\gam_a$ is the inner product and it gives
the orthogonal metric, while the antisymmetric product gives a basis bivector:
\bear
   &&\gam_a \cdot \gam_b \equiv \frac{1}{2} (\gam_a \gam_b + \gam_b \gam_a)=
   g_{ab} ~~~~~(\mbox{orthogonal metric}) \nonumber \\
   &&\gam_a \wg \gam_b \equiv \frac{1}{2} (\gam_a \gam_b - \gam_b \gam_a)
   ~~~~~(\mbox{bivector})
\lbl{1a}
\ear
In {\it symplectic Clifford algebras}, the antisymmetric product of two basis
vectors $q_a$ is the inner product and it gives the symplectic metric,
whilst the symmetric product gives a basis bivector:
\bear
   &&q_a \wg q_b \equiv \frac{1}{2} (q_a q_b - q_b q_a)=
   J_{ab}  ~~~~~(\mbox{symplectic metric}) \nonumber \\
   &&q_a \cdot q_b \equiv \frac{1}{2} (q_a q_b + q_b q_a)~~~~~(\mbox{bivector})
\lbl{1b}
\ear

The generators of an orthogonal Clifford algebra can be transformed into a basis
(the Witt basis) in which they behave as fermionic creation and annihilation
operators.
The generators of a symplectic Clifford algebra behave as bosonic creation and
annihilation operators. We will show how both kinds of operators can be united
into a single structure so that they form a basis of a `superspace'. We will
consider an action for a point particle in such superspace. It contains
the well known spinning particle action as a particular case\,\ci{Townsend}.
After quantization
we obtain, in particular, the Dirac equation. Spinors can be
constructed in terms of the orthogonal basis vectors rewritten in the
Witt basis. So a spinor of each minimal left ideal of an orthogonal Clifford
algebra is an element of the Fock space, whose basis elements are products
of creation operators acting on a vacuum which in turn is the product
of all annihilation operators
\,\ci{SpinorFock}--\ci{PavsicInverse}. The role of creation and
annihilation
operators together with the corresponding vacuum can be interchanged,
and so we obtain\,\ci{PavsicInverse} as many different vacua---and thus
different kinds of
spinors---as there are different minimal left ideals of the Clifford algebra.
This property has been exploited in order to explain the `internal' degrees
of freedom behind gauge theories of fundamental
forces\,\ci{HestenesGauge,PavsicKaluza,PavsicKaluzaLong,PavsicInverse}.
Other approaches to the unification by Clifford algebras have been investigated
in Refs.\,\ci{CliffUnific}--\ci{PavsicE8}.

Instead of  finite dimensional spaces, we can consider infinite dimensional
spaces. Then we have  description a field theory in terms of fermionic and
bosonic creation and annihilation operators.
The latter operators can be considered as being related to the basis vectors
of the corresponding infinite dimensional space.

Our approach provides a fresh view on quantization. `Quantization' is
replacement of the phase space coordinates with the corresponding basis
vectors. The latter vectors are quantum mechanical operators. We point out that,
in the symplectic case, the Poisson brackets between phase space coordinates
are equal to the wedge products (i.e., to $\frac{1}{2}$ times the commutators)
of the corresponding basis vectors. We show that, for quadratic Hamiltonians,
the latter vectors satisfy
the Heisenberg equations of motion, under the assumption that the classical
trajectories in phase space are arbitrary, and not necessarily solutions
to the classical equations of motion. The analogous holds in the orthogonal
case.

Such novel insight on the the basis vectors, namely that they give the metric
of spacetime (a subspace of a phase space), and at the same time
they have the role of quantum mechanical operators from which one
can create, e.g., spinors, could be exploited in the development
of quantum gravity.

In Sec.\,2 we review spaces with orthogonal and symplectic forms and their
role in quantization. In Sec.\,3 we present a relation between
the classical equations of motion and the Heisenberg equations for operators.
We consider the latter relation for a generic `Hamiltonian' that
is quadratic in coordinates and momenta. In Sec.\,4 we formulate a
theory of a point particle in the space of commuting and anticommuting
(Grassmann) coordinates, the action being a generalization of the
spinning particle action\,\ci{Townsend} and of the local Sp(2) symmetric
action considered in Refs.\,\ci{Bars}. In Sec.\,5 we discuss the representation
of spinors in terms of Grassmann coordinates. In Sec.\,6 we present
a theory of quantized bosonic and fermionic fields in terms
of symplectic and orthogonal Clifford algebras. We show that
there exist many possible definitions of vacua and fermionic creation
and annihilation operators, and how this fact could be exploited
for resolution of the cosmological constant problem. Finally, in Sec.\,7
we discuss what are the prospects of such theory for grand unification
and quantum gravity.

\section{Spaces with orthogonal an symplectic forms}
\subsection{The inner product and metric}

 {\it I. Orthogonal case}

The inner product of vectors $a$ and $b$ is given by
\be
(a,b)_g  = (a^a  \gamma _a  ,b^b  \gamma _b  )_g  = 
a^a  (\gamma _a  ,\gamma _b  )_g b^b   
= a^a  g_{^{a b } } b^b .
\lbl{1.1}
\ee
The metric is given by the inner product of two basis vectors: 
\be
(\gamma _a  ,\gamma _b  )_g  = g_{a b } ,
\lbl{1.2}
\ee
where $g_{a b }$ is a symmetric tensor.
For basis vectors we can take the generators of the
orthogonal Clifford algebra satisfying
\be
(\gamma _a  ,\gamma _b  )_g = \gamma _a  \cdot\gamma _b
 \equiv \frac{1}{2}(\gamma _a  \gamma _b   + \gamma _b  \gamma _a  ) 
   = g_{a b } .
\lbl{1.3}
\ee
Here vectors are Clifford numbers.
The inner product of the vectors $a$ and $b$ is given by the symmetric part
of the Clifford product
\be
(a,b)_g  = \frac{1}{2}(ab + ba) = a\cdot b .
\lbl{1.4}
\ee

 {\it II.Symplectic case}

The inner product of vectors $z$ and $z'$ is given by
\be
(z,z')_J  = (z^a q_a ,z'^b q_b )_J  = z^a (q_a ,q_b )_J z'^b  
= z^a J_{ab} z'^b ,
\lbl{1.5}
\ee
where
\be
(q_a ,q_b )_J  = J_{ab}
\lbl{1.6}
\ee
is the symplectic metric.

For a symplectic basis we can take generators of the
symplectic  Clifford algebra:
\be
(q_a ,q_b )_J = q_a  \wedge q_b \equiv \frac{1}{2}(q_a q_b  - q_b q_a ) 
  = J_{ab} 
\lbl{1.7}
\ee
Vectors are now symplectic Clifford numbers. 
The inner product of symplectic vectors $z$ and $z'$ is given by the
antisymmetric product
\be
(z,z')_J  = \frac{1}{2}(z\,z' - z'z) = z \wedge z'  .
\lbl{1.8}
\ee

Dimension of the symplectic vector space is even. Physically it is
realized as the {\it phase space}. We can split a
symplectic vector according to
\be
 z = z^a q_a  = x^\mu  q_\mu ^{(x)}  + p^\mu  q_\mu ^{(p)}~,~~~\mu=1,...,n .
\lbl{1.9}
\ee
Then we have
\be
 (z,z')_J=z^a J_{ab} z'^b=(x^\mu  p'^\nu   - p^\nu  x'^\nu  )g_{\mu \nu },
\lbl{1.10}
\ee
with
\be
J_{ab}  = \left( \begin{array}{l}
 \,\,\,0\,\,\,\,\,\,\,g_{\mu \nu }  \\ 
  - g_{\mu \nu } \,\,\,0 \\ 
 \end{array} \right) ,
\lbl{1.11}
\ee
where, depending on signature, the $n\times n$ block $g_{\mu \nu}$ is
the euclidean, $g_{\mu \nu} = \delta_{\mu \nu}$, or the Minkowski
metric, $g_{\mu \nu}=\eta_{\mu \nu}$.

Relations
\be
\frac{1}{2}[q_a ,q_b ] = J_{ab} 
\lbl{1.12}
\ee
give
\be
 \frac{1}{2}[q_\mu ^{(x)} ,q_\nu ^{(x)} ] 
= 0\,,\,\,\,\,\,\,\,\,\,\,\frac{1}{2}[q_\mu ^{(p)} ,q_\nu ^{(p)} ] = 0,
\lbl{1.13}
\ee
\be 
 \frac{1}{2}[q_\mu ^{(x)} ,q_\nu ^{(p)} ] = g_{\mu \nu }  ,
\lbl{1.14}
\ee
which are the Heisenberg commutation relations.

\subsection{Poisson bracket}

In {\it symplectic case}, the Poisson bracket betweem functions
$f(z)$ and $g(z)$ of phase space coordinates $z^a$ is given by
\be
 \{ f,g\} _{{\rm{PB}}}  \equiv \frac{{\partial f}}{{\partial z^a }}J^{ab}
  \frac{{\partial g}}{{\partial z^b }} ,
\lbl{1.15}
\ee
where
\be
 J^{ab}  = \left( \begin{array}{l}
 \,0\,\,\,\,\,\, - g^{\mu \nu }  \\ 
 g^{\mu \nu } \,\,\,\,\,\,0 \\ 
 \end{array} \right)
\lbl{1.16}
\ee
is the inverse symplectic metric.

By introducing the symplectic basis vectors,
we can rewrite the above expression as the wedge product
$\frac{\p f}{\p z^a} q^a \wg \frac{\p g}{\p z^b} q^b$. i.e.,
\be
\{ f,g\} _{{\rm{PB}}}=\frac{1}{2}\,[\,\frac{{\partial f}}{{\partial z^a }}q^a ,
\frac{{\partial g}}{{\partial z^b }}q^b ]\, 
= \,\frac{{\partial f}}{{\partial z^a }}J^{ab} \frac{{\partial g}}{{\partial z^b }}
\lbl{1.17}
\ee
If we take
\be
f = z^c \,,\,\,\,\,\,\,\,\,\,g = z^d 
\lbl{1.18}
\ee
we obtain
\be
\{ z^c,z^d\} _{{\rm{PB}}}=\frac{1}{2}[q^c ,q^d ] = J^{cd} 
\lbl{1.19}
\ee
These are the Heisenberg commutation relations for `operators' $q^c,~q^d$.

Usually by 'quantization' we mean the replacement of the classical phase
space coordinates $z^a=(x^\mu, p^\mu)$ by operators 
${\hat z}^a=({\hat x}^\mu, {\hat p}^\mu)$ that satisfy the Heisenberg
commutation relations. The above derivation reveals that the quantum mechanical
operators ${\hat z}^a =({\hat x}^\mu, {\hat p}^\mu)$ are in fact the symplectic
basis vectors $q^a$. This is true up to the factor $i$ in front of the
`momentum' part $q^{\mu(p)}$, a factor that is necessary in order to have Hermitian
momentum operator ${\hat p}^\mu$.
We also see that the Poison bracket of the phase space coordinates
is equal (apart from factor $\frac{1}{2}$) to the commutator of the corresponding
basis vectors. 

In {\it orthogonal case}, the Poisson bracket between functions
$f(\lambda)$ and $g(\lambda)$ of phase space coordinates $\lambda^a$
is given by
\be
\{ f,g\} _{{\rm{PB}}}  \equiv \frac{{\partial f}}{{\partial \lambda^a }}g^{ab}
 \frac{{\partial g}}{{\partial \lambda^b }} ,
\lbl{1.20}
\ee
where
\be
g^{ab}  = \left( \begin{array}{l}
 \,g^{\mu \nu } \,\,\,\,\,\,0 \\ 
 0\,\,\,\,\,\,\,\,\,\,\,g^{\mu \nu}   \\ 
 \end{array} \right)
\lbl{1.21}
\ee
is the inverse orthogonal metric. In Sec.\,4 we point out that in a phase space
with orthogonal metric the coordinates $\lambda^a$ must be Grassmann valued.

By introducing the  basis vectors we can rewrite the above expression as
the dot product $\left ( \frac{\p f}{\p \lambda^a} \gam^a \right )
 \cdot \left (\frac{\p g}{\p \lambda^b} \gam^b \right )$,
i.e.,
\be
\lbrace f,g \rbrace_{\rm PB}=
\frac{1}{2}\,\left\{ {\,\frac{{\partial f}}{{\partial \lambda^a }}\gamma ^a ,
\frac{{\partial g}}{{\partial \lambda^b }}\gamma ^b } \right\}\, = 
\,\frac{{\partial f}}{{\partial \lambda^a }}g^{ab} 
\frac{{\partial g}}{{\partial \lambda^b }}
\lbl{1.22}
\ee

If we take
\be
f = \lambda^c \,,\,\,\,\,\,\,\,\,\,g = \lambda^d ,
\lbl{1.23}
\ee
then we obtain
\be
\{ \lambda^c, \lambda^d\} _{{\rm{PB}}}=\frac{1}{2}\{ \gamma ^c ,\gamma ^d \}  
\equiv \frac{1}{2}(\gamma ^c \gamma ^d  + \gamma ^d \gamma ^c ) = g^{cd} 
\lbl{1.24}
\ee
These are the anticommutation relations for `operators' $\gam^c,~\gam^d$.
The generators $\gam^a$ of an orthogonal Clifford algebra are thus
`quantized' $\lambda^a$.

\subsection{Representation of Clifford numbers}

I. {\it Orthogonal Clifford algebra $Cl(2n)$}

In even dimensions we can split the generators $\gam_a$ according to
\be
\gamma _a  = (\gamma _\mu  ,\bar \gamma _\mu  )\,,\,\,\,\,\,\,\,\,\,
\mu  = 1,2,...,n
\lbl{1.25}
\ee
We can introduce the Witt basis:
 \be
\begin{array}{l}
 \theta _\mu   = \frac{1}{{\sqrt 2 }}(\gamma _\mu   + i\,\bar \gamma _\mu  ) \\
 \\ 
 \bar \theta _\mu   = \frac{1}{{\sqrt 2 }}(\gamma _\mu  
  - i\,\bar \gamma _\mu  ) \\ 
 \end{array}
\lbl{1.26}
\ee
Then the anticommutation relations (\ref{1.3}) become
\be
\theta _\mu  \cdot\bar \theta _\nu   \equiv 
\mbox{$\frac{1}{2}$} (\theta _\mu  \bar \theta _\nu   
+ \bar \theta _\nu  \theta _\mu  ) = \eta _{\mu \nu } 
\,,\,\,\,\,\,\,\,\,\,\,\theta _\mu  \cdot\theta _\nu   = 0
\,,\,\,\,\,\,\,\,\,\,\bar \theta _\mu  \cdot\bar \theta _\nu   = 0.
\lbl{1.27}
\ee
The generators $\gam_\mu,~\bar \gam_\mu,~\theta_\mu,~ \bar \theta_\mu$
can be represented:
 
 1)  as  matrices,
  
 2)  in terms of Grassmann coordinates:
\be
\theta ^\mu   \to \sqrt 2 \xi ^\mu  \,,\,\,\,\,\,\,\,\,\,
\bar \theta _\mu   \to \sqrt 2 \frac{\partial }{{\partial \xi ^\mu  }}
\lbl{1.28}
\ee
\be
\xi ^\mu  \xi ^\nu   + \xi ^\nu  \xi ^\mu   = 0.
\lbl{1.29}
\ee

II. {\it Symplectic Clifford algebra $Cl_S(2n)$}

The generators are:
\be
q_a  = (q_\mu ^{(x)} ,q_\mu ^{(p)} )\,,\,\,\,\,\,\,\,\,\,\mu  = 1,2,...,n
\lbl{1.30}
\ee
Commutation relation (\ref{1.12}) or (\ref{1.13}),(\ref{1.14}) can be
represented:
 
 1)  as 4 x 4  matrices, in which case the operators cannot
      be cast into the Hermitian form,

 2)  in terms of commuting coordinates:
\be
q_{}^{\mu (x)}  \to \sqrt 2 x^\mu  \,,\,\,\,\,\,\,\,q_\mu ^{(p)} \,\, \to \sqrt 2 \frac{\partial }{{\partial x^\mu  }}
\lbl{1.31}
\ee
\be
x^\mu  x^\nu   - x^\nu  x^\mu   = 0
\lbl{1.32}
\ee

\section{Heisenberg equations as equations of motion for basis vectors}
\subsection{General considerations}

Let us now consider a particle whose motion is described by the
following phase space action:
\be
I = \frac{1}{2}\int {d\tau \,(\dot z^a J_{ab} z^b \, + z^a K_{ab} z^b } ),
\lbl{2.1}
\ee
where
\be
\frac{1}{2}z^a K_{ab} z^b =H.
\lbl{2.2}
\ee
Here $K_{ab}$ is a symmetric $2n \times 2n$ matrix.
By varying the latter action with respect to $z^a$ we obtain the equations of
motion
\be
\dot z^a  = J^{ab} \frac{{\partial H}}{{\partial z^b }} .
\lbl{2.3}
\ee

Let us consider trajectories $z^a (\tau)$ as components of an
infinite dimensional vector
\be
z= z^{a(\tau )} q_{a(\tau )} \, \equiv \int {d\tau \,z^a (\tau )} \,q_a (\tau ),
\lbl{2.4}
\ee
where $q_{a(\tau)}$ are basis vectors satisfying
\be
 q_{a(\tau )}  \wedge q_{b(\tau ')}  = J_{a(\tau )b(\tau ')}
  = J_{ab} \,\delta (\tau  - \tau ') , 
\lbl{2.5}
\ee
and write the action (\ref{2.1}) in the form
\be
I = \frac{1}{2} \dot z^{a(\tau)} J_{a(\tau) b(\tau')} z^{b(\tau')} \, 
   + \frac{1}{2}z^{a(\tau)} K_{a(\tau) b(\tau')} z^{b(\tau'} ).
\lbl{2.6}
\ee
We have introduced
\be
\frac{1}{2} \int \dd \tau \, z^a (\tau) K_{ab} z^b (\tau) \equiv
\frac{1}{2} z^{a(\tau )} K_{a(\tau )b(\tau ')} z^{b(\tau ')} \equiv \cal{H},
\lbl{2.7}
\ee
where
\be
K_{a(\tau) b(\tau')} = K_{b(\tau') a(\tau) } = K_{ab} \delta (\tau - \tau') .
\lbl{2.7a}
\ee
The action (\ref{2.6}) gives the following form of the equations of motion:
\be
\dot z^{a(\tau )} = J^{a(\tau )b(\tau ')} 
\frac{{\partial {\cal{H}}}}{{\partial z^{b(\tau ')} }} 
= J^{a(\tau)c(\tau'')} \,K_{c(\tau'' )b(\tau ')} z^{b(\tau ')} ,
\lbl{2.8}
\ee
where $\p/\p z^{a(\tau)} \equiv \delta/\delta z^a (\tau)$ denotes
functional derivative.
Multiplying both sides of the latter equation by the basis vectors $q_{a(\tau)}$
we obtain an equivalent equation
\be
\dot z^{a(\tau )} q_{a(\tau)} 
= -q^{a(\tau)} K_{a(\tau)b(\tau ')} z^{b(\tau')} .
\lbl{2.9}
\ee
Let us now use the following relation:
\be
\dot z^{a(\tau )} q_{a(\tau)} \equiv \int \dd \tau \dot z^a (\tau) q_a (\tau)
= - \int \dd \tau z^a (\tau) \dot q_a (\tau)
 \equiv -z^{a(\tau )} \dot q_{a(\tau)},
\lbl{2.10}
\ee
that holds, if the boundary term vanishes, which we will assume is the case.
Then Eq.\,(\ref{2.9}) becomes
\be
  z^{b(\tau' )} \dot q_{b(\tau')} 
  = q^{a(\tau)} K_{a(\tau)b(\tau')} z^{b(\tau')} .
\lbl{2.11}
\ee
Eq.\,(\ref{2.11}) is equivalent to the classical equations of motion
derived from the action (\ref{2.1}). It holds for a classical trajectory
$z^{b(\tau')} \equiv z^b (\tau')$ satisfying the minimal action principle.

Let us now explore what happens if we drop the requirement that Eq.\,(\ref{2.11})
must hold for a trajectory that satisfies the minimal action principle
associated with (\ref{2.1}), and make 
 the assumption that
Eq.\,(\ref{2.11}) is satisfied for an arbitrary trajectory (path) $z^{b(\tau')}$.  
Then we have
\be
   \dot q_{b(\tau')} 
  = q^{a(\tau)} K_{a(\tau)b(\tau')} ,
\lbl{2.12}
\ee
which is an equation of motion for the basis vectors (operators) $q_{a(\tau)}
\equiv q_a (\tau)$. From Eqs.\,(\ref{2.7a}) and (\ref{2.12}) we have
\be
   \dot q_b (\tau ) = q^a (\tau )\,K_{ab} .
\lbl{2.13}
\ee
The right hand side of the latter equation can be written as 
\be
K_{ab} \, q^b = [q_a ,\hat H] , ~~~~~~\hat H = \frac{1}{2}q^a K_{ab} \, q^b
\lbl{2.14}
\ee
i.e., as the inner product (up to the factor 2) of a symplectic vector
$q^a$ with the symplectic bivector $\hat H$. Remember that
in the symplectic case the inner product is given by the commutator. 
So we obtain
\be
    \dot q_a  = [q_a ,\hat H] ,
\lbl{2.14a}
\ee
which are the Heisenberg equations of motion. 

We have found that the basis vectors of phase space satisfy the Heisenberg
equations of motion. This is in agreement with the finding of Sec.\,2 that
the quantum mechanical operators $\hat z^a =(\hat x^\mu, \hat p^\mu)$ are
in fact the symplectic basis vectors $q^a$. We have arrived
at the Heisenberg equations (\ref{2.14a}) from Eq.\,(\ref{2.11}) in which we
assumed that $z^{b(\tau)} \equiv z^b (\tau)$ was arbitrary. Arbitrary
$z^b (\tau)=(x^\mu (\tau),p^\mu (\tau))$ means that coordinates and momenta
are undetermined. In other words, if in the classical equations of motion
(\ref{2.11}) we assume that coordinates and momenta are undetermined,
then we obtain the operator equations (\ref{2.14a}). This sheds new light
on `quantization'.

\subsection{Particular physical cases}

The general form of the action (\ref{2.1}) contains particular cases
that depend on choice of $K_{ab}$ and the interpretation of coordinates
$x^\mu$ and parameter $\tau$.

\ \ (i) One possibility is to interpret $x^\mu$, $\mu =1,2,...,n$,
as coordinates of a non relativistic point particle, and $\tau$ as time $t$,
the signature being $(++++...)$
Then (\ref{2.1}) describes a non relativistic harmonic oscillator in
$n$-dimensions.

\ (ii) Another possibility is to interpret $x^\mu$, $\mu=0,1,2,...,n-1$,
as coordinates in $n$-dimensional spacetime with signature $(+---...)$,
$\tau$ as an arbitrary
monotonically increasing parameter along a particle's worldline, take
\be
   K_{ab} = \begin{pmatrix} ~{\bf 0}_{n\times n} &   0\cr
                          0&   \lambda (\tau) \eta_{\mu \nu} \end{pmatrix} ,
\lbl{2.15}
\ee
and assume that $\lambda$ is a Lagrange multiplier.
Then we have a phase space action for a massless relativistic particle
in a higher dimensional spacetime:
\be
   I[x^\mu,p_\mu] = \mbox{$\frac{1}{2}$} \int \dd \tau \, (
   {\dot x}^\mu \eta_{\mu \nu} p^\nu - {x}^\mu \eta_{\mu \nu} {\dot p}^\nu
   -\lambda p^\mu \eta_{\mu \nu} p^\nu)
   =  \int \dd \tau \, \left (p_\mu {\dot x}^\mu -
   \frac{\lambda}{2} \, p_\mu p^\mu \right )
\lbl{2.15a}
\ee
In a 4-dimensional subspace with signature $(+---)$, such particle behaves
as a massive relativistic particle.

 More generally, we can take 
\be
      {K^a}_b \equiv {K^{i\mu}}_{j \nu} = {A^i}_j {\delta^\mu}_\nu
                 = \begin{pmatrix}{A^1}_1&  {A^1}_2 \cr
                   {A^2}_1 &  {A^2}_2\end{pmatrix}  {\delta^\mu}_\nu \, , 
\lbl{2.16}
\ee
and consider ${A^i}_j (\tau)$ as Lagrange multipliers
that give three independent constraints. Then (\ref{2.1})
becomes the Bars action\,\ci{Bars}.
Functions ${A^i}_j (\tau)$ have
the role of compensating gauge field that render
the action (\ref{2.1}) invariant under local symplectic transformations
of Sp(2). Here indices $i,j=1,2$, occurring in double indices
$a\equiv i \mu,~b\equiv j \nu$, distinguish  $x^\mu$ from $p^\mu$.
Some more explanation can be found in Sec.\,4.3, where it is also shown
that the Bars action is a special case of a super phase space action.

\section{Point particle in `superspace'}
\subsection{A generalized space spanned over an orthogonal and symplectic basis}

Let us introduce the generalized vector space
whose elements are:

\be
Z = z^A q_A 
\lbl{3.1}
\ee
where
\be
 z^A  = (z^a ,\lambda ^a )\,,\,\,\,\,\,\,\,\,\,\,\,\,z^a  
 = (x^\mu  ,\bar x^\mu  )\,,\,\,\,\,\,\lambda ^a 
  = (\lambda ^\mu  ,\bar \lambda ^\mu  )
\lbl{3.2a}
\ee
\be 
 q_A  = (q_a ,\gamma _a )\,,\,\,\,\,\,\,\,\,\,\,\,\,\,q_a  
 = (q_\mu  ,\bar q_\mu  )\,,\,\,\,\,\,\,\,\gamma _a  
 = (\gamma _\mu  ,\bar \gamma _\mu  ) 
\lbl{3.2b}
\ee

The scalar part of the product of two such basis elements gives the metric

\be
\langle q_A q_B \rangle _S  = G_{AB}  = \left( \begin{array}{l}
 J_{ab} \,\,\,0 \\ 
 0\,\,\,\,\,\,g_{ab}  \\ 
 \end{array} \right)
\lbl{3.3}
\ee
where
\be
 q_a  \wedge q_b  = J_{ab}  = \left( \begin{array}{l}
 \,\,\,0\,\,\,\,\,\,\,\,\eta _{\mu \nu }  \\ 
  - \eta _{\mu \nu } \,\,\,0 \\ 
 \end{array} \right) \lbl{3.4}
 \ee
\be
 \gamma _a \cdot\gamma _b  = g_{ab}  = \left( \begin{array}{l}
 \,\eta _{\mu \nu } \,\,\,\,0 \\ 
 \,0\,\,\,\,\,\,\,\,\,\eta _{\mu \nu }  \\ 
 \end{array} \right) \lbl{3.5}\\ 
\ee

Let us consider a particle moving
in such space. Its worldline is given by:

\be
z^A  = Z^A (\tau )
\lbl{3.6}
\ee
where $\tau$ is a parameter on the worldline.

\subsection{Examples of possible actions}

A possible action is

\be
I = \,\,\,\frac{1}{2}\int {d\tau } \,\langle \dot Z^A q_A q_B \dot 
Z^B \rangle _S  = \frac{1}{2}\int {d\tau } \,\dot Z^A G_{AB} \dot Z^B 
\lbl{3.7}
\ee
Inserting the metric ({\ref{3.3}) and the coordinates (\ref{3.2a}), we have
\be
I = \frac{1}{2}\int {d\tau \,(\dot z^a J_{ab} \dot z^b  
+ \dot \lambda ^a g_{ab} } \dot \lambda ^b )
\lbl{3.8}
\ee
Since $J_{ab}=-J_{ba}$ and $g_{ab}=g_{ba}$, the first term differs from zero,
if $z^a$ are Grassmann (anticommuting) coordinates, whilst the
second term differs from zero if $\lambda^a$ are commuting coordinates.

Another possible action, more precisely, a part of a phase space action, is
\be
I = \frac{1}{2}\int {d\tau } \,\langle \dot Z^A q_A q_B  Z^B \rangle _S  
= \frac{1}{2}\int {d\tau } \,\dot Z^A G_{AB}  Z^B 
\lbl{3.9}
\ee
or
\be
I = \frac{1}{2}\int {d\tau \,\langle \dot z^a q_a q_b  z^b  
+ \dot \lambda ^a \gam_a \gam_b }  \lambda ^b  \rangle_S
= \frac{1}{2}\int {d\tau \, (\dot z^a J_{ab}  z^b  
+ \dot \lambda^a g_{ab} }  \lambda^b )
\lbl{3.10}
\ee
Here, in order to have non vanishing terms, $z^a$ must be commuting,
and $\lambda^a$ Grassmann (anticommuting) coordinates.

The corresponding canonical momenta are
\be
 p_a^{(z)}  = \frac{{\partial L}}{{\partial \dot z^a }} 
 = \frac{1}{2}J_{ab} z^b ,
\lbl{3.11}
\ee
\be
 p_a^{(\lambda )}  = \frac{{\partial L}}{{\partial \dot \lambda ^a }} 
 = \frac{1}{2}g_{ab} \lambda ^a
\lbl{3.12}
\ee 

Inserting expressions (\ref{3.4}) and (\ref{3.5}) for $J_{ab}$ and $g_{ab}$,
the action (\ref{3.10}) becomes
\be
I = \frac{1}{2}\int {d\tau \,(\dot x^\mu  \eta _{\mu \nu } \bar x^\nu  }  
   - {\dot {\bar x}}^\mu  \eta _{\mu \nu } x^\nu   
    + \dot \lambda ^\mu  \eta _{\mu \nu } \lambda ^\nu   
    + {\dot {\bar \lambda}}^\mu  \eta _{\mu \nu } \bar \lambda ^\nu  )
\lbl{3.13}
\ee
where
\be
    [x^\mu , x^\nu ] = 0,~~~~~[\bar x^\mu , \bar x^\nu ] = 0,
\lbl{3.14}
\ee    
\be
    \{\lambda^\mu , \lambda^\nu \} = 0,~~~~~
    \{\bar \lambda^\mu , \bar \lambda^\nu \} = 0 .
\lbl{3.14a}
\ee
The canonical momenta now read
\be
 p_\mu ^{(x)}  = \frac{{\partial L}}{{\partial \dot x^\mu  }} 
 = \frac{1}{2}\eta _{\mu \nu } \bar x^\nu  ,\,\,\,\,\,\,\,\,\,p_\mu ^{(\bar x)}
   = \frac{{\partial L}}{{\partial \dot \bar x^\mu  }} =
     - \frac{1}{2}\eta _{\mu \nu } x^\nu
\lbl{3.15}
\ee
\be     
 p_\mu ^{(\lambda )}  = \frac{{\partial L}}{{\partial \dot \lambda ^\mu  }} 
 = \frac{1}{2}\eta _{\mu \nu } \lambda ^\nu  ,\,\,\,\,\,\,\,
 \,p_\mu ^{(\bar \lambda )}  
 = \,\frac{{\partial L}}{{\partial \dot \bar \lambda ^\mu  }} 
 = \frac{1}{2}\eta _{\mu \nu } \bar \lambda ^\nu 
\lbl{3.16}
\ee

Upon quantization, the coordinates and momenta become operators,
\be
x^\mu  ,\,\,p_\mu ^{(x)} \,\, \to \hat x^\mu  ,\,\,\hat p_\mu ^{(x)} 
\lbl{3.17}
\ee
\be
\lambda ^\mu  ,\,\,p_\mu ^{(\lambda )} \, \to \hat \lambda ^\mu  ,\,\,
\hat p_\mu ^{(\lambda )} 
\lbl{3.18}
\ee
satisfying
\be
[\hat x^\mu  ,\hat p_\nu ^{(x)} ] 
 = i\,\delta ^\mu  _\nu  \, ,~~~~~~ 
 \,[\hat x^\mu  ,\hat x^\nu  ] = 0\,,\,\,\,\,\,
  [\hat p_\mu ^{(x)} ,\hat p_\nu ^{(x)} ] = 0, 
\lbl{3.19}
\ee
\be
 \{ \hat \lambda ^\mu  ,\hat p_\nu ^{(\lambda )} \}  
 = i\,\delta ^\mu  _\nu  \,,\,\,\,\,\
 \,\,\,\,\,\{ \hat \lambda ^\mu  ,\hat  \lambda ^\nu  \}  
 = 0\,,\,\,\,\,\,\,\,\,\,\{ \hat p_\mu ^{(\lambda )} ,
 \hat p_\nu ^{(\lambda )} \}  = 0 
\lbl{3.20}
\ee

Similar relations hold for barred quantities. Altogether,  we have
\be
 z^a \,,\,\,\,p_a^{(z)}  \to \hat z^a \,,\,\,\,\hat p_a^{(z)}
\lbl{3.21}
\ee
\be 
 \lambda ^a ,\,\,p_a^{(\lambda )}  \to \hat \lambda ^a ,\,\,
 \hat p_a^{(\lambda )} 
 \lbl{3.22}
 \ee
where the operators satisfy
\be
     [\hat z^a, \hat p_b^{(z)}] = i{\delta^a}_b
\lbl{3.23}
\ee
\be
   \{\hat \lambda^a , \hat p_a^{(\lambda )}\}=i{\delta^a}_b
\lbl{3.24}
\ee

If we insert
\be
 \hat p_a^{(z)}  = \frac{1}{2}J_{ab} \hat z^b , ~~~~ 
 \hat p_a^{(\lambda )}  = \frac{1}{2}g_{ab} \hat \lambda ^a , 
\lbl{3.25}
\ee
we obtain
\be
 \frac{1}{2}[\hat z^a \,,\,\hat z^b ] = i\,J^{ab} ,~~~~  
 \frac{1}{2}\{ \hat \lambda ^a ,\,\hat \lambda ^b \,\}  = i\,g^{ab} . 
\lbl{3.26}
\ee
But we see that the above operator equations are just
the relations for the basis vectors of the orthogonal
and symplectic Clifford algebras, provided that we
identify:
\be
 \hat z^a  = (q^\mu  ,i\bar q^\mu  ) ,~~~~
 \hat \lambda ^a  = (\gamma ^\mu  ,i\bar \gamma ^\mu  )
\lbl{3.27}
\ee

We se that `quantization' is in fact the replacements
of the coordinates $z^a ,\,\,\lambda ^a$
with the corresponding basis vectors. The only difference is in
the factor $i$  in front of $\bar q_\mu$.
Basis vectors,  entering the action, are `quantum operators',
apart from the $i$ in the relations (\ref{3.27}).

Instead of coordinates $\lambda^a$ and basis vectors $\gam_a$ it is
convenient to introduce new coordinates and new basis vectors
\be
\begin{array}{l}
 \lambda'{\,^a}  \equiv \xi^a = (\xi^\mu  ,\bar \xi^\mu  )\,,\,\,\,\,\,~~
 \,\xi ^\mu   = \frac{1}{{\sqrt 2 }}(\lambda ^\mu 
   - i\,\bar \lambda ^\mu  ) \\
   \\ 
 ~~~~~~~~~~~~~~~~~~~~~~~~~~~~~~~\bar \xi ^\mu  
  = \frac{1}{{\sqrt 2 }}(\lambda ^\mu   + i\,\bar \lambda ^\mu  ) \\ 
 \end{array}
\lbl{3.28}
\ee
\be
\begin{array}{l}
 \gamma '_a  \equiv \theta_a = (\theta_\mu  ,\bar \theta_\mu  )\,\,,\,\,\,\,\,\,\,~~
 \theta _\mu   
 = \frac{1}{{\sqrt 2 }}(\gamma _\mu   + i\,\bar \gamma _\mu  ) \\ 
 \\
~~~~~~~~~~~~~~~~~~~~~~~~~~~~~~~\bar \theta _\mu   
 = \frac{1}{{\sqrt 2 }}(\gamma _\mu   - i\,\bar \gamma _\mu  ) \\ 
 \end{array}
\lbl{3.29}
\ee
In the new coordinates we have
\be
g'_{ab}  = \gamma '_a \cdot\gamma '_b  \equiv \theta_a \cdot \theta_b
= \left( \begin{array}{l}
 0\,\,\,\,\,\,\eta _{\mu \nu }  \\ 
 \eta _{\mu \nu } \,\,0 \\ 
 \end{array} \right)
\lbl{3.30}
\ee
and the action (\ref{3.10}) becomes
\be
I = \frac{1}{2}\int d\tau \, (\dot z^a J_{ab}  z^b  
+ \dot \xi^a g'_{ab}  \xi^b )
  =\frac{1}{2}\int d\tau \, (\dot z^a J_{ab}  z^b 
   + \dot \xi^\mu \eta_{\mu \nu}  {\bar \xi}^\nu
   +{\dot {\bar \xi}}^{\,\mu} \eta_{\mu \nu} {\xi}^\nu )
\lbl{3.31}
\ee
Then the canonical momenta are
\be
 p_\mu ^{(x)}  = \frac{{\partial L}}{{\partial \dot x^\mu  }}
  = \frac{1}{2}\eta _{\mu \nu } \bar x^\nu  ,\,\,\,\,\,\,
  \,\,\,p_\mu ^{(\bar x)}  = \frac{{\partial L}}{{\partial \dot {\bar x}^\mu  }} 
  =  - \frac{1}{2}\eta _{\mu \nu } x^\nu 
\lbl{3.32}
\ee
\be  
 p_\mu ^{(\xi )}  = \frac{{\partial L}}{{\partial {\dot \xi}^\mu  }}
  = \frac{1}{2}\eta _{\mu \nu } \bar \xi ^\nu  ,\,\,\,\,\,\,\,
  \,p_\mu ^{(\bar \xi )} 
   = \,\frac{{\partial L}}{{\partial \dot {\bar \xi}^\mu  }} 
   = \frac{1}{2}\eta _{\mu \nu } \xi ^\nu  
\lbl{3.33|}
\ee
We see that now the canonically conjugate variables are
($\xi^\mu,\,\frac{1}{2}{\bar \xi}_\mu$) and 
($\bar \xi^\mu,\,\frac{1}{2}{\xi}_\mu$). In old coordinates and basis vectors
the situation was somewhat unfortunate, because the canonically conjugate
variables were ($\lambda^\mu$, $\frac{1}{2}\lambda_\mu$) and
($\bar \lambda^\mu,\,\frac{1}{2} \bar \lambda_\mu$), i.e., the canonical
momenta were essentially the same as the variables which they were
conjugated to.

The commuting coordinates $z^a=(x^\mu,{\bar x}^\mu)\equiv (x^\mu,p^\mu)$
and the symplectic
basis vectors $q_a=(q_\mu,{\bar q}_\mu)$ span a subspace that we will\
call {\it the bosonic subspace}. The Grassmann (anticommuting) coordinates
$\lambda^a=(\lambda^\mu,{\bar \lambda}^\mu)$ and the orthogonal basis
vectors $\gam_a=(\gam_\mu,{\bar \gam}_\mu)$, or equivalently, the Grassman
coordinates $\xi^a=(\xi^\mu,{\bar \xi}^\mu)$
and the Witt basis vectors $\theta_a=(\theta_\mu,{\bar \theta}_\mu)$,
span a subspace that we will call {\it the fermionic subspace}.
In Sec.\,2.3 we pointed out that
$(q^\mu,{\bar q}^\mu)$ can be represented by $(x^\mu, \p/\p x^\mu)$,
whilst $(\theta^\mu,{\bar \theta}_\mu)$ can be represented by
$(\xi^\mu,\p/\xi^\mu)$.

\subsection{Completing the super phase space action}

The action considered in the previous subsection is not complete.
An additional term is needed. Let the $\tau$-derivative in the action
\be
I = \frac{1}{2}\int {d\tau \,\dot Z^A G_{A B}  Z^B } 
\lbl{3.34}
\ee
be replaced with the covariant derivative
\be
\dot Z^A  \to \dot Z^A  + {A^A}_C Z^C ,
\lbl{3.35}
\ee
where ${A^B}_C$ depend on $\tau$.
So we obtain
\be
I = \frac{1}{2}\int {d\tau \,(\dot Z^A  + {A^A}_C Z^C) G_{A B} Z^B } 
\lbl{3.36}
\ee
which is invariant under $\tau$-dependet, i.e., {\it local}, `rotations'
of $Z^A$, and where ${A^A}_C (\tau)$ are the corresponding compensating gauge
fields. We will take into account only particular local `rotations' of $Z^A$,
namely those that rotate between the canonically conjugate pairs of the
(super) phase space variables $Z^A$. In the bosonic subspace the latter local
`rotations' manifest themselves as local symplectic transformations of Sp(2).
Then the corresponding compensating gauge fields are
${A^{i \mu}}_{j \nu}={A^i}_j {\delta^\mu}_\nu$,
where the $i,j=1,2$ distinguishes $x^\mu$ from ${\bar x}^\mu$, since
$z^a$ can be written in the form $z^a=z^{i \mu}\equiv (x^\mu, {\bar x}^\mu)$.
There are only three independent gauge fields ${A^i}_j (\tau)$; they
represent a choice of gauge, as discussed in refs\,\ci{Bars}, and have thus
the role of Lagrange multipliers. 
Analogous considerations hold in the fermionic subspace.
No kinetic term for the gauge fields ${A^A}_C$ so designed is necessary.

The action (\ref{3.36}) is the Bars action\,\ci{Bars}, generalized to the
superspace (see also Ref.\,\ci{PavsicBars}). In particular, the extra term
gives
\be
    {A^A}_C Z^C G_{A B} Z^B
    =\alpha p_\mu  p^\mu  \, + \,\beta \,\lambda ^\mu  p_\mu 
    + \gam {\bar \lambda}^\mu p_\mu,
\lbl{3.37}
\ee
where $\alpha,~\beta,~\gam$ are Lagrange multipliers contained in ${A^A}_C$.
Other choices of Lagrange multipliers ${A^A}_C$ are possible, and they
give expressions that are different from Eq.\,(\ref{3.37}). For the
bosonic subspace, this was discussed in Refs.\,\ci{Bars}.

For the case (\ref{3.37}), the action (\ref{3.36}) gives  the following
constraints:
\be
   p_\mu  p^\mu =0\, , ~~~~~~~~~\lambda^\mu  p_\mu= 0\, ,
    ~~~~~~~~~ {\bar \lambda}^\mu  p_\mu= 0 .
\lbl{3.38}
\ee
Mass comes from extra dimensions. It was shown by Bars\,\ci{Bars}
that the consistency
of the constraints  resulting from the action (\ref{3.36}) requires
one extra time like and one extra space like dimension of the bosonic subspace.
Consequently, also the dimensionality of the fermionic subspace is adequately
enlarged. So in this theory there are in fact extra
dimensions that allow for the occurrence of mass in four dimensions.

Upon quantization, the classical constraints $\lambda ^\mu  p_\mu= 0$ 
and ${\bar \lambda}^\mu  p_\mu= 0$ become two copies of 
the Dirac equation
\be
    \hat {\bar \lambda}^\mu  \hat p_\mu  \Psi = 0 , ~~~~~ {\rm and}~~~~~
    {\hat \lambda}^\mu  \hat p_\mu  \Psi = 0
\lbl{3.39}
\ee
where  
\be
   \hat \lambda ^\mu   = \gamma ^\mu \, , ~~~~~{\rm and}~~~~~  
   \hat {\bar \lambda} ^\mu   = i \bar \gamma ^\mu.
\lbl{3.40}
\ee
The state $\Psi$ can be represented:        

 1) as a column  $\hs{.7cm} \psi^\alpha (x)$,
 
 2) as a function $\hs{2cm} \psi (x^\mu  ,\xi ^\mu  )$.
\newline The ${\hat \lambda}^\mu = \gam^\mu$ can be represented
can be represented

1) as matrices  $\hs{.8cm} {(\gam^\mu)^\alpha}_\beta$,

2) as  $\hs{4cm} \xi _\mu  \, + \frac{\partial }{{\partial \xi ^\mu  }}$.
\newline We also have  $\hat {\bar \lambda}^\mu   = i\,\bar \gamma ^\mu$,  
which can be represented

 1) as matrices $\hs{.8cm} i{(\hat {\bar \gam}^\mu)^\alpha}_\beta$,
 
 2) as  $\hs{3.9cm} i\,(\xi _\mu  \, - \frac{\partial }{{\partial \xi ^\mu  }})$.
\newline where $\xi^\mu$ are Grassmann coordinates (see Eq.(\ref{3.28})).

\subsection{Building up spinors from basis vectors}

We have seen that upon quantization the classical Grassmann coordinates
$\lambda^a = (\lambda^\mu, \bar \lambda^\mu)$ become operators
$\hat \lambda^a = (\hat \lambda^\mu, \hat {\bar \lambda}^\mu)
=(\gam^\mu,~i \bar \gam^\mu)$, where $\gam^\mu,~ \bar \gam^\mu$
are generators (basis vectors) of $Cl(2n)$, or in general of $Cl(p,q),~p+q=2n$. In the
Witt basis (\ref{1.26}) the basis vectors satisfy the fermionic
anticommutation relations (\ref{1.27}). Using the basis vectors $\theta_\mu,
\bar \theta_\mu$ we can build up spinors by taking a `vacuum'
\be
\Omega  = \prod\limits_\mu  {\bar \theta} _\mu   
~~~~~~~\mbox{that satisfies}~~~~~~~~~~\bar \theta _\mu  \,\Omega  = 0
\lbl{3.41}
\ee
and acting on it by the `creation' operators $\theta_\mu$.
So we obtain a `Fock space' basis for spinors, that contains $2^n$
independent elements:
\be
s_{\alpha} \,\, = \,\,({\bf{1}}\Omega ,\,\,\,\,\,
\theta _\mu  \Omega \,,\,\,\,\theta _\mu  \theta _\nu  \Omega 
\,,\,\,\,\theta _\mu \theta_\nu \theta_\rho  \Omega \,,\,\,\,
\theta _\mu \theta_\nu \theta_\rho \theta_\sigma  \Omega, ... )
\lbl{3.42}
\ee
in terms of which any state can be expanded:
\be
\Psi_\Omega  = \sum\limits_{}^{} {\psi ^{\alpha} } s_{\alpha} 
\lbl{3.43}
\ee
With operators  $\theta _\mu  ,\,\,\bar \theta _\mu$,  
defined above, we can construct 
spinors as the elements of a minimal left ideal of $Cl(p,q),~p+q=2n$,
where $n$ is dimension of spacetime. Notice that in the case of 4-dimensional
spacetime, i.e., when $n=4$, the symplectic or bosonic phase space
is 8-dimensional. Therefore, if with every commuting phase space coordinate
$z^a=(x^\mu,{\bar x}^\mu)$ one associates a corresponding Grassmann
phase space coordinate $\lambda^a = (\lambda^\mu, {\bar \lambda}^\mu)$
(i.e., $\xi^a=(\xi^\mu, \bar \xi^\mu)$, in the Witt basis), then also
the `orthogonal' or fermionic phase space is 8-dimensional. In the usual
theory of the spinning particle, only four Grassmann coordinates $\lambda^\mu$
(or $\xi^\mu$) are considered, which then leads to the spinors of $Cl(1,3)$.
The usual theory thus imposes an asymmetry between the bosonic phase
space, which has eight dimensions, and the fermionic phase space, which
has four dimensions.

Definition (\ref{3.41}) is just one of the possible definitions of vaccuum.
In general, vacucum can be defined as the product of $n$ factors,
some of which are $\theta_\mu$ and some are $\bar \theta_\mu$:
\be
 \Omega = \theta_{\mu_1} \theta_{\mu_2} ...\theta_{\mu_r}
        \bar \theta_{\mu_{r+1}} \bar \theta_{\mu_{r+2}} ...\bar \theta_{\mu_n}
    ~,~~~~~ r=0,1,2,...,n
\lbl{3.44}
\ee  
The number of different vacua, $2^n$, is equal to the number of left minimal
ideals of $Cl(2n)$ (more precisely, $Cl(p,q),~p+q=2n$). With each vacuum
so constructed, we can associate a different Fock space basis for spinors
of the corresponding left ideal. The direct sum of all those different
spinor spaces is the Clifford algebra $Cl(p,q),~p+q=2n$, whose generic
element, denoted $\Psi$, can be expanded according to
\be
      \Psi = \psi^{\alpha \beta} s_{\alpha \beta} \equiv \psi^{\tl A} s_{\tl A}.
\lbl{3.44a}
\ee
Here the first index, $\alpha$, denotes the spinor components of a given ideal,
which is denoted by the second index, $\beta$. It is convenient to denote
the double index $\alpha \beta$ by a single index ${\tl A}$.
In the case, when we start from the 4-dimensional Minkowski spacetime with
signature $(1,3)$, the signature of phase space is $(2,6)$, and we obtain
$Cl(2,6)$.

With the Witt basis (\ref{3.29}), we obtain the spinor spaces as subspaces
of $Cl(2,6)$, but those spinor spaces do not contain the ordinary spinors
of $M_{1,3}$. The latter spinors are constructed\,\ci{Winberg,PavsicInverse}
in terms of a different Witt basis, which contains the elements
\be
\begin{array}{l} 
      \theta_1 = \frac{1}{{\sqrt 2 }}(\gamma_0   +  \gamma_3  ),~~~~~~
      \theta_2 = \frac{1}{{\sqrt 2 }}(\gamma_1  + i\,\gamma_2  ) \\ 
 \\
      \bar \theta_1 = \frac{1}{{\sqrt 2 }}(\gamma_0 -  \gamma_3  ),~~~~~~
      \bar \theta_2 = \frac{1}{{\sqrt 2 }}(\gamma_1  - i\gamma_2  ). 
 \end{array}
\lbl{3.45}
\ee
Then the full basis of the 8-dimensional phase space contains, in addition
to the above elements, also the following elements:
\be
\begin{array}{l} 
      \theta_3 = \frac{1}{{\sqrt 2 }}(\bar \gamma_0   + \bar \gamma_3  ),~~~~~~
      \theta_4 = \frac{1}{{\sqrt 2 }}(\bar \gamma_1  + i\,\bar \gamma_2  ) \\ 
 \\
      \bar \theta_3 = \frac{1}{{\sqrt 2 }}(\bar \gamma_0 - \bar \gamma_3  ),~~~~~~
      \bar \theta_4 = \frac{1}{{\sqrt 2 }}(\bar \gamma_1  - i\,\bar \gamma_2  ). 
 \end{array}
\lbl{3.46}
\ee

Taking the scalar product of $\Psi$ with the basis elements, we obtain
the components:
\be
   \langle {s^{\tl A}}^\ddg \Psi \rangle_S = \psi^{\tl A},
\lbl{3.47}
\ee
where $\langle ~ \rangle_S$ denotes the scalar part of a Clifford algebra
valued object. Operation $\ddg$ reverses the order of vectors in a
Clifford product and performs complex conjugation. More details can
be found in Ref.\,\ci{PavsicKaluzaLong}.

Components $\psi^{\tl A}$ represent a generic element $\Psi$ of the
Clifford algebra. If we take
the scalar product of $\Psi$ with basis elements of one left ideal, say,
$\alpha 1 \equiv \alpha$,
\be
   \langle {s^{\alpha}}^\ddg \Psi \rangle_S = \psi^{\alpha},
\lbl{3.47a}
\ee
then the components $\psi^{\alpha}$ represent a spinor state of this
chosen left ideal.

A state $\Psi$ can be represented by its components $\psi^{\tl A}$,
which are projections of $\Psi$ onto the basis elements. Alternatively,
since it is a Clifford algebra valued object, it can be represented by
matrices
\be
    \langle {s^{\tl A}}^\ddg \Psi \, s_{\tl B}\rangle_S 
    = {(\Psi)^{\tl A}}_{\tl B}.
\lbl{3.48}
\ee
These are $2^{2n} \times 2^{2n}$ matrices, and they belong to a reducible
representation of the Clifford algebra\footnote{
For the sake of simplicity, we will use the simplified notation $Cl(2n)$ for
$Cl(p,q),~p+q=2n$.}
 $Cl(2n)$. If instead of the
full basis $s_{\tl A}\equiv s_{\alpha \beta}$, we take the basis elements
of a chosen left ideal, e.g., $s_{\alpha 1} \equiv s_\alpha$, then
we have
\be
  \langle {s^{\alpha}}^\ddg \Psi \, s_{\beta}\rangle_S 
  = {(\Psi)^{\alpha}}_{\beta}
 =  \psi^{\tl A} \langle s^{\alpha} s_{\tl A}\, s_{\beta} \rangle_S
 = \psi^{\tl A} {(s_{\tl A})^{\alpha}}_{\beta} . 
\lbl{3.49}
\ee
This shows how every element of a Clifford algebra can be represented
as an irreducible $2^n \times 2^n$ matrix. For instance, $\gam_\mu$
can be represented by a matrix ${(\gam_\mu)^\alpha}_\beta$, and $\theta_\mu$
by ${(\theta_\mu)^\alpha}_\beta$.

\section{On the representation of spinors in terms of the Grassmann coordinates}

Operators $\theta _\mu  ,\,\,\bar \theta _\mu$, which are in fact
nothing but the generator of the Clifford algebra, expressed in the
Witt basis (\ref{1.26}) or (\ref{3.29}), can be represented either
as matrices or in terms of the Grassmann coordinates and their derivatives:
\be
\theta ^\mu   \to \sqrt 2 \xi ^\mu  \,,\,\,\,\,\,\,\,\,\,\bar \theta _\mu 
  \to \sqrt 2 \frac{\partial }{{\partial \xi ^\mu  }}
\lbl{5.1}
\ee
A spinor state is then represented by a wave function $\psi(\xi^\mu)$.
Since $\xi^\mu$ are Grassmann, anticummuting, coordinates, the Taylor
expansion of $\psi(\xi^\mu)$ has a finite number of terms, namely $2^n$,
which is the same as the number of components of a spinor in an $2n$-dimensional
space. 

The definition of vacuum $\bar \theta_\mu \Omega = 0,~\mu=1,2,...,n$,
is now represented by the equation
\be
     \frac{\partial }{{\partial \xi ^\mu  }} \Omega(\xi^\mu)=0
\lbl{5.2}
\ee
whose solution is a constant, e.g.,  $\Omega (\xi^\mu) = 1$.
A state (\ref{3.43}) can then be represented as
\be
    \Psi_\Omega \rightarrow \psi(\xi^\mu) 
  = \sum_{r=0}^{r=n} \psi_{\mu_1 \mu_2 ...\mu_r} 
  \xi^{\mu_1} \xi^{\mu_2}...\xi^{\mu_r} \, .
\lbl{5.3}
\ee
Some other definition of vacuum, e.g., 
$\theta_1 \theta_2 \bar \theta_3 \bar \theta_4 ....
\bar \theta_n \Omega =0$, is represented by the equation
\be
   \xi^1 \xi^2 \frac{\p}{\p \xi^3} \frac{\p}{\p \xi^4}....\frac{\p}{\p \xi^n}
   \Omega (\xi^\mu) = 0,
\lbl{5.4}
\ee
which has for a solution $\Omega(\xi^\mu) =\xi^1 \xi^2$. The corresponding
state is then represented as
\be
    \Psi_\Omega \rightarrow \psi(\xi^\mu) 
  = \left ( \sum_{r=0}^{r=n} \psi_{\mu_1 \mu_2 ...\mu_r} 
  \xi^{\mu_1} \xi^{\mu_2}...\xi^{\mu_r} \right ) \xi^1 \xi^2 \, .
\lbl{5.5}
\ee
In general,
a vacuum state is given by (\ref{3.44}), which can be represented by
function
\be
     \Omega(\xi^\mu) = \xi^{\mu_1} \xi^{\mu_2} ...\xi^{\mu_s} ,
\lbl{5.6}
\ee
and the corresponding spinor state is represented as
\be
    \Psi_\Omega \rightarrow \psi(\xi^\mu) 
  = \left ( \sum_{r=0}^{r=n} \psi_{\mu_1 \mu_2 ...\mu_r} 
  \xi^{\mu_1} \xi^{\mu_2}...\xi^{\mu_r} \right )
  \xi^{\nu_1} \xi^{\nu_2} ...\xi^{\nu_s}  \, .
\lbl{5.7}
\ee
The spinor states of every minimal left ideal of $Cl(p,q),~p+q=n$,
can be thus represented in terms of a function of $n$ Grassmann
coordinates $\xi^\mu$, $\mu=1,2,...,n$.

So far we have considered functions of Grassmann coordinates $\xi^\mu$.
Those functions can be considered as components of vectors (states)
$\Psi_\Omega$
with respect to a basis, say $h(\xi)$:
\be
     \Psi_\Omega = \int \dd \xi^1 \dd \xi^2 ... \dd \xi^n \psi (\xi) h(\xi).
\lbl{5.8}
\ee
This can be written in a more compact notation as
\be
    \Psi_\Omega = \psi^{(\xi}) h_{(\xi)} ,
\lbl{5.9}
\ee 
where $\psi^{(\xi)} \equiv \psi(\xi) \equiv \psi(\xi^\mu)$, and
$h_{(\xi)}\equiv h(\xi) \equiv h(\xi^\mu)$. Here $(\xi)$ is written
as an index denoting a component of the vector $\Psi_\Omega$, and if the index
is repeated, then the integration over $\xi$ in the sense of Eq.\,(\ref{5.8})
is implied. Components $\psi^{(\xi)}$ can be complex valued; then also the
basis vectors are complex (i.e., consisting of two real components, as
described in more detail later).

Basis vectors are assumed to satisfy
\be
   h^{(\xi)}.h_{(\xi')} = \mbox{$\frac{1}{2}$}
   (h^{(\xi)} h_{(\xi')}+h_{(\xi')} h^{(\xi)}) ={\delta^{(\xi)}}_{(\xi')}
\lbl{5.10}
\ee
where ${\delta^{(\xi)}}_{(\xi')}\equiv \delta (\xi-\xi')$ is the delta
`function' in the Grassmann space, satisfying 
\be
   {\delta^{(\xi)}}_{(\xi')} f^{(\xi')} = f^{(\xi)}
\lbl{5.11}
\ee
Indices are lowered and raised, respectively, by the metric
$\rho_{(\xi)(\xi')} = h_{(\xi)}.h_{(\xi')}$ and its inverse
$\rho^{(\xi)(\xi')}$. If $h^{(\xi)}$ are complex, then $h^{(\xi)}$
actually means $h^{i(\xi)}=(h^{1(\xi)},h^{2(\xi)})$ or 
$h^{i(\xi)}=(h^{~(\xi)},h^{*(\xi)})$.
Analogously, $h_{(\xi)}$ means $h_{i(\xi)}=(h_{1(\xi)},h_{2(\xi)})$ or 
$h_{i(\xi)}=(h_{~(\xi)},h_{*(\xi)})$. Metric $\rho_{(\xi)(\xi')}$ thus has an
implicit index $i=1,2$.

In Eq.\,(\ref{5.9}) we can perform the expansion in terms of $\xi^\mu$, and
use the properties $\int \dd \xi =0,~~\int \dd \xi \, \xi =1$,   which
leads to
\bear
  \psi^{(\xi}) h_{(\xi)} &=& \int \dd^n \xi \left (\psi\biggl\vert_0 +
  \frac{\p \psi}{\p \xi^\mu}\biggl\vert_0 \xi^\mu
   + \frac{\p^2 \psi}{\p \xi^\mu \p \xi^\nu}\biggl\vert_0\xi^\mu \xi^\nu + ... 
+\frac{\p^n \psi}{\p \xi^{\mu_1}... \p \xi^{\mu_n}}\biggl\vert_0 \xi^{\mu_1}
... \xi^{\mu_n} \right ) \nonumber \\
    & &  \hs{1cm}\times \left (h\biggl\vert_0 +
  \frac{\p h}{\p \xi^\mu}\biggl\vert_0 \xi^\mu
   + \frac{\p^2 h}{\p \xi^\mu \p \xi^\nu}\biggl\vert_0 \xi^\mu \xi^\nu + ... 
+\frac{\p^n h}{\p \xi^{\mu_1}... \p \xi^{\mu_n}}\biggl\vert_0 \xi^{\mu_1}
... \xi^{\mu_n}
  \right ) \nonumber\\
  &=& \psi\biggl\vert_0 \frac{\p^n h}{\p \xi^1 ... \xi^n}\biggl\vert_0 
  + \frac{\p \psi}{\p \xi^\mu}\biggl\vert_0\,
  \frac{\p^{n-1} h}{\p \xi^1 ... \hat{\p \xi^\mu} ... \p \xi^n}\biggl\vert_0 + ...
  + \frac{\p^n \psi}{\p \xi^1 ... \xi^n}\biggl\vert_0 h\biggl\vert_0 
  \nonumber \\
   &=& \psi^\alpha h_\alpha , ~~~\alpha = 1,2,..., 2^n .
\lbl{5.12}
\ear 
Here $h_\alpha$ are discrete basis vectors spanning the $2^n$-dimensional
space ${\mathbb C}^{2^n}$ in which spinors of one minimal left ideal live.
The basis vectors $h_\alpha \in {\mathbb C}^{2^n}$ are used here instead of
the basis spinors $s_\alpha \in Cl(2n)$ defined in Eq.\,(\ref{3.42}).

One can get the contact with the usual language of quantum theory by
the correspondence $h_{(\xi)} = |\xi \rangle ,~h^{(\xi)} = \langle \xi |$,
and $h_\alpha = |\alpha \rangle ,~  h^\alpha = \langle \alpha |$.
Then
\be
  \Psi_\Omega = \psi^{(\xi}) h_{(\xi)}
   = \int |\xi \rangle \dd^n \xi \,\langle \xi |\Psi_\Omega \rangle
   =\sum_\alpha |\alpha \rangle \langle \alpha |\xi \rangle \dd^n \xi 
   \, \langle \xi |\Psi_\Omega \rangle
   = h_\alpha {c^\alpha}_{(\xi)} \psi^{(\xi)}
\lbl{5.13}
\ee
The transformation coefficients
${c^\alpha}_{(\xi)} \equiv \langle \alpha |\xi \rangle$ can be read from
Eq.\,(\ref{5.12}).

Let us now consider a generic element $\Psi \in Cl(2n)$.
If we project  $\Psi$ onto the basis $h^{(\xi)}$, then we obtain
the wave function of Grassmann coordinates, 
\be
   h^{(\xi)}.\Psi = h^{(\xi)}.(\psi^{(\xi')} h_{(\xi')})  =\psi^{(\xi)}.
\lbl{5.14}
\ee
This is one possible representation of a state $\Psi$ that is analogous
to (\ref{3.47a}). In fact, it is a projection of $\Psi$ onto one ideal. 
In analogy with 
Eq.\,(\ref{3.49}), we can put $\Psi$ into a sandwich between  $h^{(\xi)}$
and $h_{(\xi)}$, and consider matrices
\be
   \langle h^{(\xi)}\, \Psi \, h_{(\xi')} \rangle_S = {(\Psi)^{(\xi)}}_{(\xi')}
\lbl{5.15}
\ee

According to Eqs.\,(\ref{3.42}),(\ref{3.43}), a state $\Psi$ is expressed
in terms of basis vectors $\theta_\mu, \bar \theta_\mu$, which,
in turn can be expressed
as linear combinations of $\gam_\mu,~\bar \gam_\mu$ (generators of $Cl(2n)$).
Therefore, the following matrix elements are of particular interest:
\bear
 & &\theta_\mu \rightarrow 
  \langle h^{(\xi)}\, \theta_\mu \, h_{(\xi')} \rangle_S
  = \xi_\mu \, \delta (\xi - \xi') \nonumber \\
  & & \bar \theta_\mu \rightarrow 
    \langle h^{(\xi)}\, \bar \theta_\mu \, h_{(\xi')} \rangle_S
    = \frac{\p}{\p \xi^\mu} \, \delta (\xi - \xi) .
\lbl{5.16}
\ear
They can be used as basic blocks for building up a matrix
(\ref{5.15}) that represents a generic state $\Psi$, spanned over
a spinor basis of all $2^n$ minimal left ideals of $Cl(2n)$.

Whereas a matrix ${(\Psi)^{(\xi)}}_{(\xi')}$ can represent any element $\Psi$ of
$Cl(2n)$, components $\psi^{(\xi)}$ represent a spinor $\Psi_\Omega$
of one ideal only. From a 1st rank spinor, i.e., a spinor of one left
minimal ideal,
\be
     \Psi_\Omega = \psi^\alpha h_\alpha= \psi^{(\xi)} h_{(\xi)},
\lbl{5.17}
\ee
and its reverse\footnote{We define reversion so to include complex
conjugation `*'.},
interpreted as a spinor of a right minimal ideal
\be
     \Psi_\Omega^\ddg = \psi^{* \alpha} h_\alpha^\ddg 
     = \psi^{*(\xi)} h_{(\xi)}^\ddg,
\lbl{5.18}
\ee
we can pass to a 2nd rank spinor by taking the tensor product
\be
      \Psi_\Omega \otimes \Psi_\Omega^{'\ddg} 
      = \psi^\alpha \psi^{'*\beta} h_\alpha \otimes h_\beta^\ddg
      = \psi^{(\xi)}\psi^{'*(\xi')} h_{(\xi)} \otimes h_{(\xi')}^\ddg .
\lbl{5.19}
\ee

Once we have the bases $h_\alpha \otimes h_\beta^\ddg$ and 
$h_{(\xi)} \otimes h_{(\xi')}^\ddg$, we can span over them the space
of objects whose components are $\psi^{\alpha \beta}$ and
$\psi^{(\xi)(\xi')}$, respectively. Components $\psi^{\alpha \beta}$
represent a generic element of $Cl(2n)$. Similarly, also components
$\psi^{(\xi)(\xi')}$ represent a generic element of $Cl(2n)$.
Instead of the double indices we can use the single indices, and
write $\psi^{\alpha \beta} \equiv \psi^{\tl A}$ (as we did in
Eq.\,(\ref{3.44a})), or $\psi^{(\xi)(\xi')} \equiv \psi^{(\xi,\xi')}$.
Instead of the tensor product basis $h_\alpha \otimes h_\beta^\ddg$
we can take the basis $s_{\alpha \beta}\equiv s_{\tl A}$, used in
Eq.\,(\ref{3.44a})), and instead of  $h_{(\xi)} \otimes h_{(\xi')}^\ddg$
we can take another basis, denoted $h_{(\xi)(\xi')}\equiv h_{(\xi,\xi')}$,
which spans the space of Grassmann functions. An element of the latter
space is
\be
    \Psi = \psi^{(\xi,\xi')} h_{(\xi,\xi')}
\lbl{5.20}
\ee
Whereas  a spinor of one left ideal is described by a function
$\psi^{(\xi)} \equiv \psi (\xi)$
of $n$ Grassmann coordinates $\xi \equiv \xi^\mu,~\mu=1,2,..., n$,
a generic element of $Cl(2n)$ is described by a function
$\psi^{(\xi,\xi')} \equiv \psi (\xi,\xi')$ of $2n$
Grassmann coordinates
$(\xi,\xi') \equiv (\xi^\mu,\xi'^\mu),~\mu, \nu=1,2,..., n$.
If we perform an expansion, analogous to that of Eq.\,(\ref{5.12}),
we obtain $2^{2n}$ independent components.

In general, a wave function depends on commuting
coordinates $x^\mu$ as well, because it has to form a representation, not
only of the orthogonal basis vectors $(\gam_\mu, {\bar \gam}_\mu)$ or
$(\nth_\mu,{\bar \nth}_\mu)$, but also of the symplectic basis
vectors $(q_\mu, {\bar q}_\mu) \equiv (q_{\mu}^{(x)}, q_{\mu}^{(p)})$
according to Eq.\,(\ref{1.31}). Therefore, instead of Eq.\,(\ref{5.9}),
we have
\be
    \Psi_\Omega = \psi^{(x,\xi}) h_{(x,\xi)} ,
\lbl{5.21}
\ee 
where
\be 
     h^{(x,\xi)} \cdot h_{(x',\xi')} = {\delta^{(x,\xi)}}_{(x',\xi')} .
\lbl{5.22}
\ee
A state $\Psi_\Omega$ is now an element of an infinite dimensional space spanned
over a basis $h_{(x,\xi)}$, the components being a wave function
$\psi^{(x,\xi)} \equiv \psi (x,\xi )$. In the next section we will formulate
foundations of the field theory based on the orthogonal and symplectic
Clifford algebras.

\section{Description of fields}

\subsection{Bosonic fields}

In Sec.\,3, Eq.\,(\ref{2.1}), we considered the phase space action, $I[z^a]$,
$z^a=(x^\mu,p^\mu)$,  of a point particle. The latter action is a functional
of coordinates and momenta.
If, by the equations of motion, the momenta can be expressed in terms
of coordinates, then we can obtain an equivalent action, $I[x^\mu]$, which is
a functional of coordinates only. For instance, instead of the phase space
action (\ref{2.15a}), we obtain
\be
    I[x^\mu] = \mbox{$\frac{1}{2}$} \int \dd \tau \, 
    \frac{{\dot x}^\mu {\dot x}_\mu}{\lambda} .
\lbl{6.1}
\ee

We can interpret the coordinates $x^\mu$, $\mu=1,2,...,n$, in several
different ways, as we did in Sec.\,3.2.
For instance, the coordinates $x^\mu$ can be interpreted to denote position
(of a non relativistic
particle) in $n$-dimensional space, $\mathbb{R}^n$. The latter space
is a vector space, spanned over the set of $n$-basis vectors, $\gam_\mu$,
that can be generators of the Clifford algebra $Cl(n)$,
satisfying\footnote{Alternatively, $x^\mu$ can be
interpreted as denoting position (of a relativistic particle) 
in $n$-dimensional spacetime, $M_n\equiv \mathbb{R}^{(1,n-1)}$, with
signature $(1,n-1)$. Then, instead of $\delta_{\mu \nu}$, we would have
the Minkowski metric, $\eta_{\mu \nu}$.}
\be
     \gam_\mu \cdot \gam_\nu \equiv  \mbox{$\frac{1}{2}$}
     (\gam_\mu \gam_\nu + \gam_\nu \gam_\mu) = \delta_{\mu \nu}~,
\lbl{6.2}     
\ee
so that the action (\ref{6.1}) reads
\be
   I[x^\mu] = \mbox{$\frac{1}{2}$} \int \dd \tau \, 
   \frac{\dot x^\mu \gam_\mu \gam_\nu \dot x^\nu} {\lambda}
\lbl{6.1a}
\ee
Position in $\mathbb{R}^n$ is thus described by a vector $x=x^\mu \gamma_\mu$
of the orthogonal Clifford algebra $Cl(n)$. Instead of the
configuration space, and an action $I[x^\mu]$, such as (\ref{6.1}),
we can consider the
corresponding phase space, whose points are described by symplectic vectors
$z=z^a q_a$, considered in Sec.\,2, the action  $I[x^\mu,p^\mu]$ being, e.g.,
Eq.\,(\ref{2.15a}) or, in general, Eq.\,(\ref{2.1}).

Analogously,  we can consider a state vector $\Phi=\phi^{(x)} h_{(x)}$,
spanned over an infinite set of basis vectors $h_{(x)}$ that satisfy
the relations of a generalized orthogonal  infinite dimensional
Clifford algebra
\be
  h_{(x)} \cdot h_{(x')} = \delta_{(x)(x')} ~.
\lbl{6.2a}
\ee
 The latter vector is an infinite dimensional
analog of a vector $x=x^\mu \gam_\mu$ considered in previous paragraph.
Equations of ``motion" for $\phi (\tau,x)\equiv \phi^{(x)}(\tau)$
can be derived from an action functional
\be
I[\phi (\tau,x)] = \int \dd \tau \, {\cal L}(\phi,\p_\tau \phi, \p_\mu \phi) ,
\lbl{6.2b}
\ee 
which can be any one known in the field theory,
e.g., the scalar field action
\be
  I[\phi]=\mbox{$\frac{1}{2}$} \int \dd \tau \, \dd^n x 
  \, \left ( (\p_\tau \phi)^2 +  \p_\mu \phi \, \p^\mu \phi 
   -m^2 \phi^2 \right ).
  \lbl{6.3}
\ee
where  $x^\mu$ are coordinates of space ${\mathbb R}^n$,
$\p_\tau \equiv \p/\p\tau$, and $\p_\mu \equiv \p/\p x^\mu$.
The latter action can be written in the form
\be
   I[\phi] = \mbox{$\frac{1}{2}$} \int \dd \tau \, 
   \left ( \p_\tau \phi^{(x)} h_{(x)} h_{(x')} \p_\tau \phi^{(x')} +
   \p_\mu \phi^{(x)} h_{(x)} h_{(x')} \p^\mu \phi^{(x')} -
   m^2 \phi^{(x)} h_{(x)} h_{(x')} \phi^{(x')} \right ) ,
\lbl{6.3a}
\ee
which is an infinite dimensional analog of the action (\ref{6.1a}).

Introducing the momentum $\Pi (x) = \delta {\cal L}/\delta {\dot \phi} =
{\dot \phi}$ and the Hamiltonian 
$H=\int \dd^n x\, (\Pi(\tau,x) {\dot \phi}(\tau,x) - {\cal L})$ we obtain
the {\it phase space action}
\bear
    I[\phi,\Pi] &=& \int \dd \tau \, 
    \left [ \int \dd^n x\, \Pi {\dot \phi} - H \right ] \nonumber \\
   &=& \mbox{$\frac{1}{2}$} \int \dd \tau \, \dd^n x \, \left [
   \Pi {\dot \phi} - {\dot \Pi} \phi - (\Pi^2 - \p^\mu \phi \, \p_\mu \phi
   + m^2 \phi^2) \right ] .
\lbl{6.3b}
\ear
The above equations also  hold for complex fields, if an implicit index,
$c=1,2$, denoting, e.g., the real and imaginary component, is assumed,
with understanding
that $\phi^2 \equiv \phi^c \phi_c = \phi^1 \phi^1 + \phi^2 \phi^2$,
or $\phi^2 \equiv \phi^c \phi_c = \frac{1}{2}(\phi^* \phi + \phi \phi^*)$.

The action (\ref{6.3b}) is a particular case of a generic phase space
action for bosonic fields that we will consider in the following.

\subsubsection{A generic phase space action for bosonic fields and
its quantization}
 
Let us consider a vector $\Phi=\phi^{i(x)} k_{i(x)}$, $i,j =1,2$, in an infinite
dimensional space, and assume that the basis vectors $k_{i(x)}$ satisfy
the relations of a symplectic Clifford algebra:
\be
 k_{i(x)}\wg k_{j(x')} \equiv \mbox{$\frac{1}{2}$} 
 (k_{i(x)}k_{j(x')}-k_{j(x')}k_{i(x)})
 = J_{i(x)j(x')} .
\lbl{6.4}
\ee
The symplectic metric has the following form:
\be
J_{i(x)j(x')}  = \left( \begin{array}{l}
 \,\,\,0\,\,\,\,\,\,\,\,\,\,\,\,\delta _{(x)(x')}  \\ 
  - \delta _{(x)(x')} \,\,\,\,0 \\ 
 \end{array} \right) ,
\lbl{6.5}
\ee
where $\delta _{(x)(x')}  \equiv \delta (x - x')$.
Shortly, $J_{i(x)j(x')} = \epsilon_{ij} \delta (x-x')$, with
$\epsilon_{ji}=-\epsilon_{ij}$.

Components $\phi^{i(x)}=(\phi^{1(x)},\phi^{2(x)})\equiv (\phi^{(x)},\Pi^{(x)})$
are analogous to
$z^a=(x^\mu,p^\mu)$ of Secs. 2 and 3, i.e., to coordinates and momenta.

The symplectic vector reads explicitly
\be
   \Phi = \phi^{i(x)} k_{i(x)} = \phi^{1(x)} k_{1(x)} +\phi^{2(x)} k_{2(x)}
   = \phi^{(x)}{ k_\phi}_{(x)} + \Pi^{(x)} {k_\Pi}_{(x)}
\lbl{6.5a}
\ee

The action is now
\be
   I =\mbox{$\frac{1}{2}$} \int \dd \tau  
   \left ( {\dot \phi}^{i(x)} J_{i(x)j(x')} \phi^{j(x')} +
           \phi^{i(x)} K_{i(x)j(x')} \phi^{j(x')} \right ) ,
\lbl{6.6}
\ee
where the fields $\phi^{i(x)}$ are assumed to be functions of $\tau$,
so that $\phi^{i(x)}$ means $\phi^{i(x)} (\tau) \equiv \phi^i (\tau,x)$.
It gives the following equations of motion:
\be
    {\dot \phi}^{i(x)} = J^{i(x)j(x')} \frac{\p H}{\p \phi^{j(x')}},
\lbl{6.7}
\ee        
where $H=\phi^{i(x)} K_{i(x)j(x')} \phi^{j(x')}\equiv
\int \dd^n x \, \dd^n \, x'\phi^i (x) K_{ij}(x,x') \phi^j (x')$,
and $\p/\p \phi^{j(x')} \equiv \delta /\delta \phi^j (x')$. If we now follow
the analogous procedure as in Eqs.\,(\ref{2.6})--(\ref{2.14a}),
we arrive at the equations of motion for the operators:
\be
  {\dot k}_{j(x')} = k^{i(x)} K_{i(x)j(x')} = [k_{j(x')},{\hat H}] ,
\lbl{6.8}
\ee
where
\be
    {\hat H} = \mbox{$\frac{1}{2}$} k^{i(x)} K_{i(x)j(x')} k^{j(x')} .
\lbl{6.9}
\ee
The latter operator equation of motion can be derived from the action
\be
   I =\mbox{$\frac{1}{2}$} \int \dd \tau  
   \left ( {\dot k}^{i(x)} J_{i(x)j(x')} k^{j(x')} +
           k^{i(x)} K_{i(x)j(x')} k^{j(x')} \right ) .
\lbl{6.10}
\ee
We obtain
\bear
     {\dot k}^{i(x)} &=& \lbrace k^{i(x)}, {\hat H} \rbrace_{\rm P.B.}
     = \frac{\p k^{i(x)}}{\p k^{m(x')}} J^{m(x')n(x'')}
      \frac{\p {\hat H}}{\p k^{n(x'')}} \nonumber \\ 
      &=& J^{i(x)n(x'')}\frac{\p {\hat H}}{\p k^{n(x'')}}
      =[k^{i(x)},{\hat H}] .
\lbl{6.11}
\ear
In the above calculation we have assumed that
$\p k^{i(x)}/\p k^{m(x')}={\delta^{i(x)}}_{m(x')}$.
Indices are lowered and raised by the symplectic metric $J_{i(x)j(x')}$
and its inverse $J^{i(x)j(x')}$.

\subsubsection{Some particular cases}

The general form of the action (\ref{6.6}) or its quantum version
(\ref{6.10}) contains particular cases that depend on choice of
$K_{i(x)j(x')}$, and on the space the coordinates $x$ are associated
with. Let us consider some of the cases:

 \ \ (i) For instance, let $x\equiv x^\mu$, $\mu=1,2,3$,
be three spatial coordinates, and $\tau =t$ the non relativistic time.
Instead of $x^\mu$, we will now write $x^r,~r=1,2,3$, so to have a closer
contact with the conventional notation.

\ \ \ \ \ a) If we take
\be
   K_{i(x)j(x')} =  \begin{pmatrix} (m^2+\p^r \p_r)\delta(x-x')  &  0\cr
                          0&  \delta(x-x')  \end{pmatrix} ,
\lbl{6.11a}
\ee
then the action (\ref{6.6}) becomes the scalar field phase space action
(\ref{6.3b}).  

\ \ \ \ \ b) If we take
\be
 K_{i(x)j(x')}=  \left ( - \frac{1}{2m}\, \p^r \p_r 
    + V(x) \right ) \delta (x-x') \, g_{ij} ~,
    ~~~~g_{ij} = \begin{pmatrix} 0 & 1 \cr
                                 1 & 0  \end{pmatrix} 
\lbl{6.12}
\ee
then the action (\ref{6.6}) becomes
\be
   I = \int \dd t \, \dd^3 x  
   \left [ \mbox{$\frac{1}{2}$}\,{\dot \phi}^i (t,x) \epsilon_{ij} \phi^j (t,x)
          + i \phi^i (t,x) \left (- \frac{1}{2m} \p^r \p_r +V(x) \right )
           g_{ij}  \phi^j (t,x) \right ] ,
\lbl{6.13}
\ee
where  $\phi^i (t,x)=(\phi(t,x),\Pi(t,x)) = (\phi(t,x),i \phi^* (t,x))$. 
In Eq.\,(\ref{6.13}) we have the action for the classical
Schr\" odinger field.

Similarly, using (\ref{6.10}), the action for operators becomes
\be
   I =\int \dd t \, \dd^3 x  
   \left [ \mbox{$\frac{1}{2}$} \, {\dot k}^i (x) \epsilon_{ij} k^j (x) +
          i k^i (x) \left (- \frac{1}{2m} \p^r \p_r +V(x) \right )
           g_{ij}  k^j (x) \right ] ,
\lbl{6.14}
\ee
which is the action for the quantized Schr\" odinger field.

\ (ii) Alternatively, let $x\equiv x^\mu$, $\mu=0,1,2,3$,  be four
coordinates of spacetime, and $\tau$ a Lorentz invariant evolution
parameter. Then the choice
\be
 K_{i(x)j(x')}=  \left ( - \frac{1}{2\Lambda} \p^\mu \p_\mu 
     \right ) g_{ij} \delta (x-x') ~, ~~~~~
    g_{ij} = \begin{pmatrix} 0 & 1 \cr
                                 1 & 0  \end{pmatrix}  
\lbl{6.15}
\ee
for a constant $\Lambda$, gives the Stueckelberg action\,\ci{Stueckelberg}
\be
   I = \int \dd \tau \, \dd^4 x  
   \left [ \mbox{$\frac{1}{2}$} \,{\dot \phi}^i (x) \epsilon_{ij} \phi^j (x) +
           i \phi^i (x) \left (- \frac{1}{2\Lambda } \p^\mu \p_\mu \right )
           g_{ij}  \phi^j (x) \right ] ,
\lbl{6.16}
\ee
and the analogous action for the quantized field $k^i (x)$.

\subsection{Fermionic fields}

Let us now consider the vector $\Psi=\psi^{i(x)} h_{i(x)}$, $i=1,2$, and
assume that the basis vectors $h_{i(x)}$ satisfy the relations of an
orthogonal Clifford algebra
\be  
    h_{i(x)} \cdot h_{j(x')}  = \mbox{$\frac{1}{2}$} (h_{i(x)} h_{j(x')}
    +h_{j(x')} h_{i(x)}) = \rho_{i(x)j(x')} ,
\lbl{6.17}
\ee
where $\rho_{i(x)j(x')}$ is an orthogonal metric. Its explicit form
depends on a chosen basis. In a particular basis the
metric can be
\be
    \rho'_{i(x)j(x')} = \delta_{ij} \delta_{(x)(x')} =
    \begin{pmatrix} \delta_{(x)(x')}  &  0\cr
                          0&  \delta_{(x)(x')}  \end{pmatrix} .
\lbl{6.18}
\ee
This can be transformed into another basis, namely the Witt basis, in which
\be
    \rho_{i(x)j(x')} = 
    \begin{pmatrix}  0   &  \delta_{(x)(x')}\cr
                        \delta_{(x)(x')}  &   0 \end{pmatrix} .
\lbl{6.19}
\ee

We assume that an implicit spinor index, $\alpha=1,2,...,2^n$, occurs
in the expressions. Thus, $\Psi=\psi^{i(x)} h_{i(x)}\equiv 
\psi^{i \alpha(x)} h_{i \alpha(x)}$.
Components $\psi^{i(x)} = (\psi^{1(x)},\psi^{2(x)}) \equiv 
(\psi^{(x)},\pi^{(x)})$ are {\it Grasmann valued} phase space variables,
and are analogous to $\xi^a \equiv \xi^{i \mu} =(\xi^\mu,{\bar \xi}^\mu)$
considered in Sec.\,4.

The fermionic field action that corresponds to the bosonic field action
(\ref{6.6}) is
\be
   I =\mbox{$\frac{1}{2}$} \int \dd \tau  
   \left ( \psi^{i(x)} \rho_{i(x)j(x')} {\dot \psi^{j(x')}} +
           \psi^{i(x)} H_{i(x)j(x')} \psi^{j(x')} \right ) ,
\lbl{6.20}
\ee
If we now repeat a procedure, analogous to that in 
Eqs.\,(\ref{6.6})--(\ref{6.11}), then we obtain the following equations
of motion for the basis vectors
\be
     {\dot h}^{i(x)} = \lbrace h^{i(x)}, {\hat H} \rbrace ,
\lbl{6.21}
\ee
where ${\hat H} = h^{i(x)} H_{i(x)j(x')} h^{j(x')}$, and brace means
the anticommutator.

Let, in particular, be $x \equiv x^r,~r=1,2,3$, and $\tau = t$. If $\rho_{i(x)j(x')}$
is given by Eq.\,(\ref{6.19}), and
\be
 H_{i(x)j(x')}=
 \begin{pmatrix}  0   & -({\alpha^*}^r \p_r +i m) \delta(x-x') \cr
                ({\alpha}^r \p_r +i m) \delta(x-x')   &   0 \end{pmatrix} ,
\lbl{6.22}
\ee 
where $\alpha^r = \gam^0 \gam^r$ are hermitian matrices in the spinorial
indices, i.e., ${(\alpha^r)}_{\beta \alpha}^* = (\alpha^r)_{\alpha \beta}$,
then the action (\ref{6.20}) becomes
\be
   I=\int \dd t \, \dd^3 x \, \left [ \mbox{$\frac{1}{2}$}(\pi {\dot \psi}
   - {\dot \pi} \psi) + \pi \gam^0 \gam^r \p_r \psi - m \pi \psi \right ],
\lbl{6.23}
\ee
where we have taken into account the anticommutativity of $\psi$ and $\pi$.
If we write $\pi = i \psi^\dg$ and omit the total derivative
$(\dd/\dd t) (\pi \psi)$, we obtain the usual phase space action for the
Dirac field. The latter field can be the usual $2^{n/2}$-component Dirac
field in $n$-dimensional spacetime, or it can be a field with
$2^n$ components, considered in Secs.\,4.4 and 5.

Instead of a discrete spinor index $\alpha$ we can take the Grassmann
coordiantes $\xi$ and write the vector $\Psi$ in the form
$\Psi = \psi^{i(x,\xi)} h_{i(x,\xi)}$, where the basis vectors satisfy
\be
    h_{i(x,\xi)} \cdot h_{j(x',\xi')} = \rho_{i(x,\xi)j(x',\xi')}
    = \delta_{ij} \delta_{(x)(x')} \delta_{(\xi)(\xi')} .
\lbl{6.24}
\ee
If, as a model, we take two Grassmann coordinates only,
$\xi \equiv (\xi^1,\xi^2)$, then
$\Psi$ is a usual 4-component spinor (see Eq.\,(\ref{3.45})). A generic
fermionic field action can be written as
\be
   I =\mbox{$\frac{1}{2}$} \int \dd \tau  
   \left ( \psi^{i(x,\xi)} \rho_{i(x,\xi)j(x',\xi')} {\dot \psi^{j(x',\xi')}} +
           \psi^{i(x,\xi)} H_{i(x,\xi)j(x',\xi')} \psi^{j(x',\xi')} \right ) .
\lbl{6.25}
\ee

In particular, for $H_{i(x,\xi)j(x',\xi')}$ we can take a matrix, analogous to
(\ref{6.22}), in which objects $\gam^\mu$ are represented in terms of
$(\xi \pm \p/\p \xi) \delta (\xi-\xi')$.

We can consider, for instance, the case (\ref{3.45}), and represent 
\be
\nth^1 \rightarrow \xi^1 \delta(\xi-\xi'),
~~~\nth^2 \rightarrow \xi^2 \delta(\xi-\xi'),
~~~{\bar \nth}_1 \rightarrow \frac{\p}{\p \xi^1}\delta(\xi-\xi'),
~~~{\bar \nth}_2 \rightarrow \frac{\p}{\p \xi^2}\delta(\xi-\xi') ,
\lbl{6.26}
\ee
and invert the relations (\ref{3.45}) so to express $\gam^\mu$ in
terms of $\nth^1,~\nth^2, ~{\bar \nth}_1,~{\bar \nth}_2$.
Then the action (\ref{6.25}) is equivalent to
the usual action for the 4-component spinor field.

Alternatively, we can consider the case  (\ref{3.29}), in which the
number of vectors, spanning the fermionic phase space is the same
as the number of vectors spanning the bosonic phase space, namely eight. Then
$\xi \equiv \xi^\mu, ~\mu=0,1,2,3$, and a vector
$\Psi = \psi^{i(x,\xi}) h_{i(x,\xi)}$, for a fixed $i$, represents
a 16-component spinor field.

\subsection{Poisson brackets}

In Sec.\,2.2 we have observed that in the symplectic case the Poisson
bracket of the phase space coordinates is equal to the wedge product
(i.e., to one half times the commutator) of the corresponding symplectic
basis vectors. Similarly, in the orthogonal case, the Poisson bracket
of the phase space variables is equal to one half of the anticommutator
of the corresponding orthogonal basis vectors. Analogous holds for fields.

In {\it symplectic case} we have
\bear
\left\{ {f(\phi^{i(x)} ),g(\phi^{j(x)} )} \right\}_{{\rm{PB}}}
  &=& \frac{{\partial f}}{{\partial \phi^{i(x)} }}J^{i(x)j(x')}
   \frac{{\partial g}}{{\partial \phi^{j(x')} }} \nonumber \\
  &=& \frac{{\partial f}}{{\partial \phi^{i(x)} }}
  \mbox{$\frac{1}{2}$} [ {k^{i(x)} ,\,k^{j(x')} } ]
   \frac{{\partial g}}{{\partial \phi^{j(x')} }} 
\lbl{6.27}
\ear
In particular, if
$f(\phi^{i(x)})=\phi^{k(x'')}$ and $\,g(\phi^{j(x)})=\phi^{\ell(x''')}$, then
\be
  \{ {\phi^{k(x'')},\phi^{\ell(x''')} } \}_{{\rm{PB}}}
 = J^{k(x'')\ell (x''')}
  = \mbox{$\frac{1}{2}$} [ {k^{k(x'')} ,\,k^{\ell(x''')} } ]
\lbl{6.28}
\ee
The Poisson bracket of two classical fields is equal to the symplectic metric
which, in turn, is equal to the wedge product (i.e., $1/2$ times the commutatot)
of two symplectic basis vectors.
The latter vectors are bosonic field operators, and from Eq.\,(\ref{6.28})
we see that the canonical commutation relations are in fact
the relations (\ref{6.4}) of a symplectic Clifford algebra.

In {\it orthogonal case} we have
\bear
\left\{ {f(\psi^{i(x)} ),g(\psi^{i(x)} )} \right\}_{{\rm{PB}}}
  &=& \frac{{\partial f}}{{\partial \psi^{i(x)} }}\rho^{i(x)j(x')}
   \frac{{\partial g}}{{\partial \psi^{j(x')} }} \nonumber \\
  &=& \frac{{\partial f}}{{\partial \psi^{i(x)} }}
  \mbox{$\frac{1}{2}$} \{ {h^{i(x)} ,\,h^{j(x')} } \}
   \frac{{\partial g}}{{\partial \psi^{j(x')} }} 
\lbl{6.29}
\ear
In particular, if
$f(\psi^{i(x)})=\psi^{k(x'')}$ and $\,g(\psi^{i(x)})=\psi^{\ell(x''')}$, then
\be
  \{ {\phi^{k(x'')},\phi^{\ell(x''')} } \}_{{\rm{PB}}}
 = \rho^{k(x'')\ell (x''')}
  = \mbox{$\frac{1}{2}$} \lbrace {k^{k(x'')} ,\,k^{\ell(x''')} } \}
\lbl{6.30}
\ee
The Poisson bracket of two classical fields is equal to the orthogonal
metric. On the other hand, the orthogonal metric is equal to the
symetrized product (given by the anticommutator) of two orthogonal
basis vectors. The latter vectors are fermionic field operators, and
Eq.\,(\ref{6.30}) shows that the canonical anticommutation relations
for fermionic fields are in fact the relations (\ref{6.17}) of an
orthogonal Clifford algebra.

Spinor indices $\alpha,~\beta$ are not explicitly displayed in
Eqs.\,(\ref{6.29}),(\ref{6.30}).
Another possibility is to rewrite the latter equations by replacing
$\psi^{i(x)} \equiv \psi^{i \alpha (x)}$ with $\psi^{i(x,\xi)}$.

\subsection{Generalization to `superfields'}

Both actions (\ref{6.6}),(\ref{6.20}) can be unified into a single action
by introducing a `superfield'
\be
     {\bf \Psi} ={\psi}^A h_A
\lbl{6.31}
\ee
where $\psi^A =(\phi^{i(x)},\psi^{i(x)})$ and $h_A=(k_{i(x)},h_{i(x)})$.
So we have
\be
   I[\psi^A] = \mbox{$\frac{1}{2}$} \int \dd \tau ({\dot \psi}^A G_{AB} \psi^B
      + \psi^A H_{AB} \psi^B)
\lbl{6.32}
\ee
where
\be
    \langle h_A h_B \rangle_S = G_{AB} =
    \begin{pmatrix}  J_{i(x)j(x')} &  0 \cr
                      0 &          \rho_{i(x)j(x')} \end{pmatrix}
\lbl{6.33}
\ee
If
\be
    H_{AB} = \begin{pmatrix}    K_{i(x)j(x')} &  0 \cr
                      0 &          H_{i(x)j(x')} \end{pmatrix}
\lbl{6.34}
\ee
then the action (\ref{6.33}) is exactly the sum of the bosonic field action
(\ref{6.6}) and  the fermionic field action (\ref{6.20}).
In general, $H_{AB}$ may have non vanishing off diagonal terms which
are responsible for a coupling between the fermionic and the bosonic fields.

Again we kept the spinor indices hidden, so that $\psi^{i(x)}$
actually meant $\psi^{i\alpha (x)}$. If, instead, we take
$\psi^{i(x,\xi)}$, then the discrete spinor components arise from the
expansion of $\psi^{i(x,\xi)} \equiv \psi^i (x,\xi)$ in terms of the
Grassmann coordinates $\xi\equiv \xi^\mu$. Our superfield can then be
written as
\be
    {\bf \Psi} = \phi^{i(x)} k_{i (x)} + \psi^{i (x,\xi)} h_{i(x,\xi)}
\lbl{6.35}
\ee
But in such expression for a superfield there is an assymmetry between
the bosonic and the fermionic part. A more symmetric expression is
\be 
   {\bf \Psi} = \phi^{i(x,\xi)} k_{i(x,\xi)} + \psi^{i (x,\xi)} h_{i(x,\xi)}
\lbl{6.36}
\ee
so that both parts contains the commuting coordinates $x$ and the 
anticommuting (Grassmann) coordinates $\xi$. Then both fields, the
commuting $\phi^{i(x,\xi)}$ and the anticommuting $\psi^{i (x,\xi)}$, can
form a representation for the symplectic  basis vectors
$(q^\mu,{\bar q}_\mu)\rightarrow (x^\mu,\p/\p x^\mu)$ and for the orthogonal
basis vectors $(\nth^\mu,{\bar \nth}_\mu)\rightarrow (\xi^\mu, \p/\p \xi^\mu)$.
The action is then that of Eq.\,(\ref{6.32}) in which the matrices
$G_{AB}$ and $H_{AB}$ are given by suitably generalized Eqs.\,(\ref{6.33})
and (\ref{6.34}), in which instead of $K_{i(x)j(x')}$ and $H_{i(x)j(x')}$
we have $K_{i(x,\xi)j(x',\xi')}$ and $H_{i(x,\xi)j(x',\xi')}$, respectively.

But isn't such a theory in conflict with the connection
between spin and statistic that comes from the requirement
of microcausality? Not necessarily. Had we taken for the
submatrix $K_{i(x,\xi)j(x',\xi')}$ in Eq.\,(\ref{6.34}) a
``Dirac equation like matrix"
such as (\ref{6.22}), then the action (\ref{6.32}) would contain a part
with a bosonic field described by the Dirac action. This would be
problematic. But a scalar field like matrix (\ref{6.11a}) would pose no problem.
The fact that the bosonic field depends also on Grassmann coordinates,
$\xi^\mu$, means that we have a number of bosonic fields, coming
from the expansion of $\phi^i (x^\mu,\xi^\mu)$ in terms of $\xi^\mu$.
Each field is in agreement with microcausality, i.e., their
commutators at different times all vanish outside the light cone.

Such   `superfield' (\ref{6.36}) with the action (\ref{6.32}) seems to be
a natural generalization of the point particle in superspace considered in
Sec.\,4. A deeper investigation of this topics is beyond the scope of the
present paper.

\subsection{Fock space states}

We will now explicitly show how the basis vectors, $h_{i(x,\xi)}$ and
$k_{i(x)}$, that span, respectively, an orthogonal and a symplectic
phase space, behave as creation and annihilation operators
of a quantum field theory.

\subsubsection{Fermionic fields: Generators of orthogonal Clifford algebras}

Let 
\be
\Psi = \psi^{i(x,\xi)} h_{i(x,\xi)} = \psi^{i \alpha (x)} h_{i \alpha (x)},
~~~i=1,2,~~~\alpha =1,2,...,2^{\boldsymbol{n}},
\lbl{6.36a}
\ee
be a vector of the fermionic subspace of the
total (super) phase space, where the basis vectors satisfy
\be
  h_{i(x,\xi)} \cdot h_{j(x',\xi')} 
  = \delta_{ij} \delta_{(x)(x')} \delta_{(\xi)(\xi')},
\lbl{6.37}
\ee
or
\be
  h_{i \alpha (x)} \cdot h_{j \beta (x')} 
  = \delta_{ij} \delta_{(x)(x')} \delta_{\alpha \beta} .
\lbl{6.38}
\ee
Introducing the Witt basis,
\be
   h_{\alpha (x)} 
      = \mbox{$\frac{1}{\sqrt{2}}$} (h_{1 \alpha (x)}+i h_{2 \alpha (x)}),
\lbl{6.39}
\ee
\be
   \bar h_{\alpha (x)} 
      = \mbox{$\frac{1}{\sqrt{2}}$} (h_{1 \alpha (x)}-i h_{2 \alpha (x)}),
\lbl{6.40}
\ee      
the anticommutation relations (\ref{6.37}),(\ref{6.38}) become the
familiar relations for fermionic creation and annihilation operators:
\be
   h_{\alpha (x)} \cdot \bar h_{\beta (x')}
    = \delta_{\alpha \beta} \delta_{(x)(x')} ,
\lbl{6.41}
\ee
\be
     h_{\alpha (x)} \cdot h_{\beta (x')} =0, ~~~~ 
     \bar h_{\alpha (x)} \cdot \bar h_{\beta (x')} =0 .
\lbl{6.42}
\ee
A possible vacuum state is the product of all operators $\bar h_{\alpha (x)}$:
\be
    \Omega = \prod_{\alpha, x} \bar h_{\alpha (x)}   ,
\lbl{6.43}
\ee
\be
     \bar h_{\alpha (x)} \Omega = 0.
\lbl{6.43a}
\ee     
A basis of the Fock space is then
\be
   \lbrace  h_{\alpha_1 (x_1)} h_{\alpha_2 (x_2)}\, ...  \, h_{\alpha_r (x_r)}
   \Omega \rbrace ~, ~~~~ r=0,1,2,3,...  .
\lbl{6.44}
\ee
For a fixed point $x$ we have
 \be
   \lbrace  h_{\alpha_1 (x)} h_{\alpha_2 (x)}\, ...  \, h_{\alpha_r (x)}
   \Omega \rbrace ,~~~~~ r = 0,1,2,..., D ,
\lbl{6.45}
\ee
or, more explicitly,
\be   
    \lbrace \Omega,~ h_{\alpha_1 (x)} \Omega,~
   h_{\alpha_1 (x)} h_{\alpha_2 (x)} \Omega,\,... \, , 
   h_{\alpha_1 (x)} h_{\alpha_2 (x)} ... h_{\alpha_D (x)}\Omega \rbrace .
\lbl{6.46}
\ee    
Here $D=2^{\boldsymbol{n}}$ is the number of the basis vectors $h_{\alpha (x)}$, i.e.,
the creation operators (\ref{6.39}) that arise from expansion
of the basis vectors $h_{(x,\xi)}$ in terms of $\boldsymbol{n}$
Grassmann coordinates
$\xi \equiv \xi^{\boldsymbol{\mu}},~\boldsymbol{\mu} = 1,2,..., 
\boldsymbol{n}$. We now also consider the case in which
the number $\boldsymbol{n}$ of anticommuting (Grassmann) coordinates
is different
from the number $n$ of the commuting coordinates, $x^\mu,~\mu=1,2,...,n$.
This is distinguished by using normal and bold symbols.
In particular, $D=16$, if $\boldsymbol{n}=4$ (which was our choice), and
$D=4$, if $\boldsymbol{n}=2$ (which is the usual choice). The dimension of the Fock space
spanned over the basis (\ref{6.46}) is $2^D$. In particular,
\be
     2^D =  \left\{ \begin{array}{ll}
             2^{16} & \textrm{if  $\boldsymbol{n}=4$}\\
             2^4  &  \textrm{if $\boldsymbol{n}=2$}
             \end{array} \right.
\lbl{6.47}
\ee

Besides the choice of vacuum (\ref{6.43}), there are other possible choices,
analogous to those considered in Sec.\,4.4, for example,
\be
    \Omega = \prod_{\alpha, x}  h_{\alpha (x)}~,~~~~~~~~
     h_{\alpha (x)} \Omega = 0.
\lbl{6.43b}
\ee
More generaly, we have
\be
    \Omega = \left ( \prod_{\alpha \in R_1, \, x} \bar h_{\alpha (x)}
    \right ) \left (\prod_{\alpha\in R_2, \,x} h_{\alpha (x)}
    \right ) .
\lbl{6.48}
\ee
Here $R=R_1 \cup R_2$ is the set od indices $\alpha=1,2,...,D$,
where $R_1$ and $R_2$ are subsets of indices, e.g., $R_1 =
\{ 1,4, 7, ...,D\}$ and $R_2 =\{2,3,5,6,...D-1 \}$. Altogether,
with respect to all such arrangements of indices $\alpha$, there are
$2^D$ possible vacua, giving $2^D \times 2^D =2^{2D}$
basis states. Since we consider operators $h_{\alpha (x)}$
$\bar h_{\alpha (x)}$ at a fixed point $x$, we can factor out from $\Omega$
the part due to the product over $x$, and so those $2^{2D}$
states span a Clifford algebra $Cl(2D)$. In particular,
we have $Cl(32)$, if $\boldsymbol{n}=4$, and $Cl(8)$, if $\boldsymbol{n}=2$.
To sum up, 
a Clifford albegra $Cl(2D)$ is generated at every point $x$ by a set of
$2D$ basis vectors, $(h_{\alpha (x)}, \bar h_{\alpha (x)})$, or equivalently,
by $(h_{(x,\xi)}, \bar h_{(x,\xi)})$, which are fermionic creation and
annihilations operators.

If we do not factor out from $\Omega$ the part due to the product over
points $x \in {\mathbb R}^n$,
then it turns out that there are many other possible definitions of vacuum,
such as
\be
    \Omega = \left ( \prod_{\alpha, \, x\in {\cal R}_1} \bar h_{\alpha (x)}
    \right ) \left (\prod_{\alpha, \,x\in {\cal R}_2} h_{\alpha (x)}
    \right ) ,
\lbl{6.48a}
\ee
depending on a partitition of ${\mathbb R}^n$ into two domains
${\cal R}_1$ and ${\cal R}_2$ so that ${\mathbb R}^n = {\cal R}_1 \cup
{\cal R}_2$. Instead of the configuration space, we can take the
momentum space, and consider, e.g., positive and negative momenta $p$.
If ${\mathbb R}^n$ is the Minkowski spacetime, then we can have a
vacuum of the form
\be
    \Omega = \left ( 
        \prod_{\alpha, \, p^0>0,{\bf p}} \bar h_{\alpha (p^0,{\bf p})} \right ) 
       \left ( 
        \prod_{\alpha, \, p^0<0,\,{\bf p}}  h_{\alpha (p^0,\,{\bf p})} \right ) 
\lbl{6.49}
\ee
which is annihilated according to
\be
    \bar h_{\alpha (p^0>0,{\bf p})}\, \Omega = 0~,~~~~ 
     h_{\alpha (p^0<0,\,{\bf p})} \,\Omega = 0,
\lbl{6.50}
\ee
whereas one particle states are created according to
\be
     h_{\alpha (p^0 >0,{\bf p})}\, \Omega~,~~~~ 
     \bar h_{\alpha (p^0<0,\,{\bf p})} \,\Omega.
\lbl{6.51}
\ee
With respect to the above vacuum (\ref{6.49}), one kind of particles 
are created by positive energy unbarred operators
$h_{\alpha (p^0 >0,{\bf p})}$, whilst the other kind of particles are
created by negative energy barred operators
$~\bar h_{\alpha (p^0<0,\,{\bf p})}$. The vacuum with
reversed properties can also be defined, besides many other possible vacua.
All those  vacuum definitions participate in a description
of the interactive processes of elementary particles. What we take into account
in our current quantum field theory calculations seem to be only a part of a
larger theory that has been neglected. It could be that some of the
difficulties (e.g., infinities) that we have encountered, are partly due
to neglection of such a larger theory. For instance, the vacuum (\ref{6.50})
is considered by Jackiw et al.\,\ci{Jackiw} within the context of a
2-dimensional field theory with signature $(+-)$.
In Refs.\,\ci{PavsicPseudoHarm} it is shown how with such definition of vacuum
we can obtain vanishing zero point energy, and yet the Casimir and other
such effects remain intact. This could be a resolution\,\ci{PavsicPseudoHarm}
of the problem of the huge cosmological constant predicted by the ordinary
quantum field theory. Another application is in string theory which,
as shown in Ref.\,\ci{PavsicString}, can be formulated in non critical
dimensions

\subsubsection{Bosonic fields: Generators of symplectic Clifford
algebras}

Let us now consider a vector
\be
   \Phi = \phi^{i(x)} k_{i(x)}~,~~~~~i=1,2
\lbl{6.52}
\ee
of the bosonic subspace of the total (super) phase space, where the basis
vectors satisfy relation (\ref{6.4}) of a symplectic Clifford algebra.
In the basis
\be
{k'}_{1(x)}\equiv k_{(x)} =\mbox{$\frac{1}{\sqrt{2}}$}(k_{1(x)} + k_{2(x)}),
\lbl{6.53a}
\ee
\be
{k'}_{2(x)} \equiv \bar k_{(x)} = \mbox{$\frac{1}{\sqrt{2}}$} (k_{1(x)} - k_{2(x)})
\lbl{6.53b}
\ee
relations (\ref{6.4}) become
\be
   k_{(x)} \wg k_{(x')} = 0,~~~~~ \bar k_{(x)} \wg \bar k_{(x')} = 0,
\lbl{6.54}
\ee
\be
     \bar k_{(x)} \wg k_{(x')} = \delta_{(x)(x')},
\lbl{6.55}
\ee
which, apart from a factor $1/2$ that enters definition of the wedge
product, are the commutation relations for bosonic creation and
annihilation operators.

In the case of fermionic operators, a possible vacuum was defined as the
product of all annihilation operators. For boson operators such definition
does not work. At the moment it is not clear to me whether a bosonic
vacuum can be defined in terms of creation and annihilation operators.
Formally, we define
\be
   \bar k_{(x)} \Omega = 0,
\lbl{6.56}
\ee
where $\Omega$ now denotes a bosonic vacuum. The basis
\be
   \{ k_{(x_1)} k_{(x_2)}... k_{(x_r)} \}~,~~~~~r=0,1,2,...
\lbl{6.57}
\ee
spans a Fock space, whose vectors are
\be
   \Phi_F = \sum_{r=0}^\infty \phi^{(x_1)(x_2)...(x_r)}
    k_{(x_1)} k_{(x_2)}... k_{(x_r)} \Omega~ ,
\lbl{6.58}
\ee
where components $\phi^{(x_1)(x_2)...(x_r)}$ are symmetric in
$(x_1)(x_2)...(x_r)$.

On the other hand, a vector (\ref{6.52}) can be generilzed to an element
of a symplectic Clifford algebra:
\be
      \Phi_C = \sum_{r=0}^\infty \phi^{i_1(x_1)i_2(x_2)...i_r(x_r)}
    k_{i_1(x_1)} k_{i_2(x_2)}... k_{i_r(x_r)}  ,
\lbl{6.59}
\ee 
which contain both kinds of operators, $k'_{1(x)}\equiv k_{(x)}$ and
$k'_{2(x)}\equiv \bar k_{(x)}$. It remains to be explored whether 
the vectors $\Phi_F$ of the form (\ref{6.58}) belong to a subspace of the space
whose vectors are $\Phi_C$ of Eq.\,(\ref{6.59}). This is equivalent to
the question of whether $\Omega$ can be defined in terms of $\bar k_{(x)}$.

\section{Discussion}

\subsection{Prospects for unification}

We started from the super phase space action (\ref{3.36}) in which the number
of commuting variables $z^a (\tau) =(x^\mu (\tau),p^\mu (\tau))$ is the
same as the number of anticommuting variables
$\lambda^a (\tau)=(\lambda^\mu (\tau),\bar \lambda^\mu (\tau))$,
$\mu=1,2,...,n$. This arises, if we consider superfields
$Z^a (\tau, \zeta)$ which depend on a commuting parameter $\tau$, and
on an anticommuting parameter $\zeta$. So we have
$Z^a (\tau, \zeta) = z^a (\tau) + \zeta \lambda^a (\tau)$.

After (first) quantizatiom
we arrived at the action (\ref{6.20}) or (\ref{6.25}) for a vector field
$\Psi=\psi^{i(x,\xi)} h_{i(x,\xi)} = \psi^{i \alpha (x)} h_{i \alpha (x)}$,
$\alpha = 1,...,2^n$ and $i=1,2$. The latter index denotes a field and
its canonically conjugate field. The basis vectors $h_{i \alpha (x)}$,
satisfying relations (\ref{6.38}), are generators of an infinite
dimensional Clifford algebra, and they have the role of quantized fields.
If we transform them into another basis according to (\ref{6.39}),(\ref{6.40}),
then the new operators $h_{\alpha (x)},~\bar h_{\alpha (x)}$ satishy the
anticommutation relations (\ref{6.41}),(\ref{6.42}) for fermioinic fields.
If we start from 4-dimensional spacetime, then the index $\alpha$ assume
$2^4=16$ values. Therefore, the set of of fields
$\{ h_{\alpha (x)},~\bar h_{\alpha (x)} \}$ is bigger than it is necessary
for description of a spin $\frac{1}{2}$ particle and its antiparticle, in
which case four values of $\alpha$ are sufficient. We now have a possibility
that $\{ h_{\alpha (x)},~\bar h_{\alpha (x)} \}$ describe electron, neutrino,
and, e.g., the corresponding mirror particles, as shown in
Ref.\,\ci{PavsicInverse}.

In order to describe other particles and gauge
interactions of the standard model, one has to extend the theory.
One possibility \,\ci{PavsicKaluza,PavsicKaluzaLong,PavsicMaxwellBrane}
is to replace spacetime with Clifford
space\,\ci{CastroHint}--\ci{CastroPavsicReview}, which is a
manifold of dimension $N=2^n$, whose
tangent space at any point $X$ is a Clifford algebra $Cl(n)$. Clifford space
is a quenched configuration space associated with
$p$-branes\,\ci{QuenchedBrane,PavsicMaxwellBrane}. One can then proceed as
we did in this paper,
and arrive at the set of fermionic field operators
$\{ h_{\alpha (X)},~\bar h_{\alpha (X)} \}$, where $\alpha$ now runs
from 1 to $2^N$, because we replaced $n$-dimensional spacetime with
$N$-dimensioanl Clifford space. The theory then becomes analogous to
a unification in the presence of higher dimensions.
Another possibility is to exploit the $2^D$-dimensional Fock space spanned
over the basis (\ref{6.46}), and take into
account the fact that the latter Fock space is a left minimal ideal of
a Clifford algebra $Cl(2D)$,  and is thus the space of spinors. Various
approaches to the unification of fundamental particles and forces by
Clifford algebras have been explored in
Refs.\,\ci{HestenesGauge}--\ci{PavsicE8},\,\ci{PavsicInverse}.

\subsection{Prospects for quantum gravity}

We have seen in Sec.\, 2.3 that the generators $\gam_a = (\gam_\mu,\bar \gam_\mu)$,
$\mu =1,2,...,n$,
of an orthogonal Clifford algebra $Cl(2n)$ can be rewritten in terms of
new generators, $\theta_a = (\theta_\mu, \bar \theta_\mu)$,
which satisfy the fermionic anti commutation relations (\ref{1.27}).
In Sec.\,4.4. we then show that $\theta_\mu , \bar \theta_\mu$ act as
fermionic creation and annihilation operators, from which we can build
basis spinors, $s_{\tl A}$, of all minimal left ideals of $Cl(2n)$.
A generic element is $\Psi = \psi^{\tl A} s_{\tl A}$, where
the coefficients $\psi^{\tl A}$ may depend on spacetime position $x^\mu$.
An interesting object to consider is
\be
      \langle \gam_\mu \rangle_1 
      = \langle \Psi^\ddg (x) \gam_\mu \Psi (x) \rangle_1 \,,
\lbl{7.1}
\ee     
i.e., the vector part of the expectation
value of a vector $\gam_\mu$. Since $\Psi$ is a generic element
of Clifford algebra, a ``Clifford aggregate" or polyvector, the expectation
value is a linear superposition of vectors $\gam_{\mu'}$:
\be
    \langle \gam_\mu \rangle_1 =  {e_\mu}^{\nu'}(x) \,\gam_{\nu'} .
\lbl{7.2}
\ee
Here $\gam_{\nu'}$ are orthogonal vectors, whilst $\langle \gam_\mu \rangle_1$
need not be orthogonal, and ${e_\mu}^{\nu'}$ may serve the role of vielbein.
There is a possibility that the vectors $\langle \gam_\mu \rangle_1$ are
tangent vectors to manifold with non vanishing curvature, so that their
inner product 
\be
\langle \gam_\mu \rangle_1 \cdot \langle \gam_\nu \rangle_1 = g_{\mu \nu} (x)
\lbl{7.3}
\ee
gives a metric that cannot be transformed into
$\eta_{\mu \nu}$ at every point $x$. If it is indeed 
the case that the curvature can be different from zero, then curved
space(time) can be generated from position
dependent spinors. This remains to be explored, and if it turns to be
true, this will have implications for quantum gravity.

\section{Conclusion}

In this work we have pointed out how `quantization' can be seen from yet another
perspective. We reformulated and generalized the theory of quantized
fields.  An action for a physical system, such as a point particle or a field,
can be written in the phase space form, and it contains either the symplectic
or the orthogonal form (or both).
The corresponding basis vectors satisfy either the fermionic
anticommutation relations or the bosonic commutation relations.
If we take a Hamiltonian that is quadratic in the phase space variables,
derive the classical equations of motion and then assume that coordinates and
momenta are undetermined, it turns out that the basis vectors satisfy
the Heisenberg equations of motion. Quantum mechanical
operators are just the basis vectors included in the phase space action.
They can be expressed as creation and annihilation operators acting
on a vacuum which is the product of annihilation
operators (in the fermionic case). In the finite dimensional case this
gives the Fock basis for spinors. If we consider not only one vacuum,
but all possible vacua, then we obtain the Fock basis for a Clifford
algebra\,\ci{BudinichFock,PavsicInverse}. In infinite dimensional case,
i.e., in the case of fields, we obtain a more general Fock basis and
many possible vacua that go beyond those usually considered in quantum
field theories. It would be interesting to explore whether such a
generalized quantum field theory is free of the difficulties, such as
infinities and the cosmological constant problem.

As a particular model we considered a point particle described in terms
of commuting and an equal number, $n$, of anticommuting (Grassmann) phase
space variables. The phase space action (\ref{3.36}) contains a form that
consists of the latter variables and the corresponding basis vectors---
the generators of an orthogonal, $Cl(2n)$, and a symplectic Clifford algebra,
$Cl_S(2n)$. If we start
from a 4-dimensional spacetime, i.e., if we take $n=4$, then we obtain
the spinor states---created from the basis vectors---of sufficiently high
dimensionality that they can be considered in our attempts for grand
unification.
Finally, we showed how
the fact that the basis vectors on the one hand are quantum mechanical
operators, and on the other hand they give metric, could be exploited in the
development of quantum gravity.

\vs{3mm}

\centerline{\bf Acknowledgement}

\vs{2mm}

This work is supported by the Slovenian Research Agency.

\end{document}